\documentclass[twocolumn,aps,prb,10pt,floatfix,superscriptaddress,longbibliography]{revtex4-2}
\usepackage{graphicx}
\usepackage{dcolumn}
\usepackage{bm}
\usepackage{amsmath}
\usepackage{amssymb}
\usepackage{siunitx}
\usepackage{color}
\usepackage{hyperref}
\usepackage{physics}
\usepackage[normalem]{ulem}
\renewcommand{\vec}[1]{\mbox{\boldmath$\mathrm{#1}$}}

\begin{document}

\preprint{APS/123-QED}

\title{Effects of interlayer Dzyaloshinskii–Moriya interaction on the shape and dynamics of magnetic twin-skyrmions}

\author{Tim Matthies}
\affiliation{Department of Physics, University of Hamburg, 20355 Hamburg, Germany}
\author{Levente Rózsa}
\email{Corresponding author: rozsa.levente@wigner.hun-ren.hu}
\affiliation{Department of Theoretical Solid State Physics, Institute for Solid State Physics and Optics, HUN-REN Wigner Research Centre for Physics, H-1525 Budapest, Hungary}
\affiliation{Department of Theoretical Physics, Institute of Physics, Budapest University of Technology and Economics, M\H{u}egyetem rkp. 3, H-1111 Budapest, Hungary}
\author{Roland Wiesendanger}
\affiliation{Department of Physics, University of Hamburg, 20355 Hamburg, Germany}
\author{Elena Y. Vedmedenko}
\affiliation{Department of Physics, University of Hamburg, 20355 Hamburg, Germany}

\date{\today}

\begin{abstract}

Abstract\\
Magnetic skyrmions have been proposed as promising candidates for storing information due to their high stability and easy manipulation by spin-polarized currents. Here, we study how these properties are influenced by the interlayer Dzyaloshinskii--Moriya interaction (IL-DMI), which stabilizes twin-skyrmions in magnetic bilayers. We find that the spin configuration of the twin-skyrmion adapts to the direction of the IL-DMI by elongating or changing the helicities in the two layers. Driving the skyrmions by spin-polarized currents in the current-perpendicular-to-plane configuration, we observe significant changes either in the skyrmion velocity or in the skyrmion Hall angle depending on the current polarization. These findings unravel further prospects for skyrmion manipulation enabled by the IL-DMI.

\end{abstract}

\maketitle

 \section{\label{sec:int}Introduction}

Chirality is an intriguing physical property of key importance for modern spintronic devices. The manipulation of chiral magnetic textures has become the focus of many applications, such as magnetic memories, logic devices, neuromorphic and unconventional computing~\cite{Back2020,Grollier2020,parkin2021,Goebel2021,Marrows2024}. While most studies have focused on planar geometries, recent investigations of spin structures modulated in all three dimensions have opened new prospects for 3D spintronics. Three-dimensional spin structures have been studied in bulk chiral magnets~\cite{Azhar2022,Zheng2023}, curvilinear systems~\cite{Makarov2021}, specially shaped nanomagnets~\cite{fernandez2017three}, and magnetic multilayers~\cite{Legrand2018}. 

Skyrmions in synthetic antiferromagnets, constructed from ferromagnetic layers coupled by the interlayer exchange coupling (IEC), provide an example of such 3D structures~~\cite{Dohi2019,xia2021TubesSynthAntiFM,legrand2020room,Qiu2021SynthAntiSkyr,juge2022skyrmions,Matheus2024SkyrAntiferro}. A two-dimensional skyrmion is formed inside each layer, with the spin directions in the skyrmions being reversed between the layers due to the antiferromagnetic coupling. The effect of ferromagnetic IEC on skyrmion size and stability has been studied in Refs.~\cite{DEGER2019165399,Schrautzer2022}. Skyrmions may often be characterized by two independent parameters~\cite{Nagaosa2013}: the vorticity and the helicity. The vorticity is the integer winding number determining how many times and in which direction the in-plane spins wind around the out-of-plane spin in the center, which determines the topological charge $Q$ when also considering the direction of the center spin. 
The helicity angle characterizes the rotational sense of the spins when moving along the radial direction out from the center, including Bloch-type and N\'{e}el-type rotations. The total topological charge of synthetic antiferromagnetic skyrmions is zero due to the symmetric but opposite magnetization between the layers, which forces the skyrmions to move along the driving spin-polarized current in the current-in-plane (CIP) geometry, leading to a suppression of the skyrmion Hall effect~\cite{zhang2016magnetic}. The cancellation of the topological charge also strongly enhances the diffusion in these systems~\cite{Dohi2023}. The helicity influences the direction of skyrmion motion in the current-perpendicular-to-plane (CPP) geometry as well~\cite{Hrabec2017,Weienhofer2022Stochastic,Msiska2022}, which includes the effect of the so-called spin--orbit torque~\cite{Manchon2019SOT}. Quantum helicity eigenstates have been proposed as a basis for skyrmion qubits~\cite{psaroudaki2022}. The helicity has been demonstrated to vary in thick magnetic multilayers due to demagnetization effects~\cite{Legrand2018}, and it may also be influenced by strain and an external magnetic field~\cite{liu2024}.


The antisymmetric counterpart of the IEC is the interlayer Dzyaloshinskii–Moriya interaction (IL-DMI)~\cite{Dzyaloshinsky1958,Moriya1960}, which gathered recently a lot of attention in the research community~\cite{zhao2025interlayer}. This chiral coupling leads to a non-zero chirality between the layers~\cite{Vedmedenko2022,Han2019,Guo2022,Arregi2023,CascalesSandoval2025}, which could facilitate field-free spin-orbit torque switching of the magnetization~\cite{pai2024}. However, the influence of the IL-DMI on the equilibrium structure and the dynamics of skyrmions in multilayers remains unexplored.

Here, we demonstrate that when two magnetic layers hosting skyrmions are coupled via IL-DMI, a three-dimensional topological structure becomes stable, which we call a twin-skyrmion. We study the structure and the current-driven motion of such twin-skyrmions using atomistic spin-dynamics simulations, which are found to be in favorable agreement with semi-analytical calculations of the skyrmion profile and the analytical description of the dynamics based on a collective-coordinate approach. Our theoretical calculations show that 
the size 
and shape of magnetic twin-skyrmions can be effectively controlled by the IL-DMI, while the helicity may differ 
in the top and bottom magnetic layers. 
These changes in the skyrmion profiles are demonstrated to influence the velocity and the skyrmion Hall angle. Since the IL-DMI can be manipulated by electric fields in synthetic magnetic bilayers~\cite{Kammerbauer2023}, 
twin-skyrmions may be promising candidates for spintronic applications based on three-dimensional spin structures. 

\begin{figure*}
    \centering
    \includegraphics[width=0.9\linewidth]{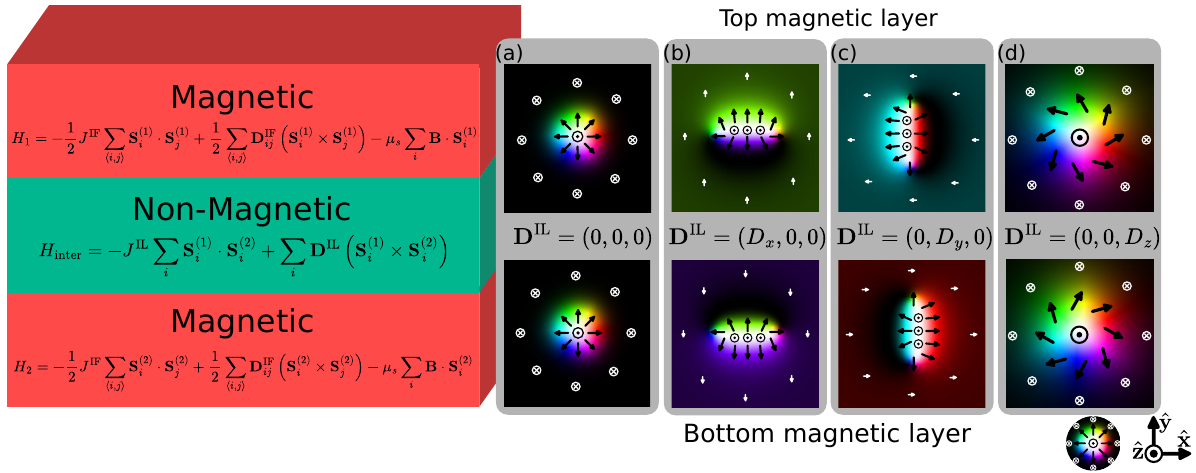}
    \caption{Effect of IL-DMI on Néel-type twin-skyrmions. A schematic of the physical system studied here is shown on the left, consisting of 
    two magnetic layers and a non-magnetic layer between them. The contributions to the Hamiltonian in Eq.~\eqref{eq:HamiltonianSplit} from the different layers are written inside them. 
    Without IL-DMI the material can host typical Néel-type skyrmions, as seen in (a). If the IL-DMI lies in the $xy$ plane, the twin-skyrmion becomes 
    elongated along the direction of the IL-DMI vector, as seen in (b) and (c). Finally, if the IL-DMI points into the out-of-plane $z$ direction, 
    the skyrmions in the two layers twist relative to each other by changing their helicities. The parameters are $J^{\text{IF}}=\SI{12}{meV}$, $|\vec{D}^\text{IF}|=\SI{3}{meV}$, $|\vec{J}^\text{IL}|=\SI{0.6}{meV}$, 
    $B^{z}=\SI{-3.5}{T}$. The IL-DMI is $\vec{D}^{\textrm{IL}}=\vec{0}$ for (a), $\vec{D}^{\textrm{IL}}=\SI{8}{meV}\cdot \hat{\vec{x}}$ for (b), $\vec{D}^{\textrm{IL}}=\SI{8}{meV}\cdot \hat{\vec{y}}$ for (c), and $\vec{D}^{\textrm{IL}}=\SI{1}{meV} \cdot \hat{\vec{z}}$ for (d). 
    }
    \label{fig:schematic}
\end{figure*}

\section{\label{sec:results}Results}
\subsection{Considered system}
We describe the magnetic system by a classical atomistic spin model of two layers of a square lattice containing $128\times 128$ sites on top of each other. We split 
the 
spin Hamiltonian of our system into three parts,
\begin{align}
    H=H_1+H_2+H_{\text{inter}},
    \label{eq:HamiltonianSplit}
\end{align}
where $H_1$ includes all 
interactions inside the first layer, $H_2$ the interactions in the second layer, and $H_{\text{inter}}$ includes all interlayer interactions. Denoting a spin in a layer with the unit vector $\vec{S}_i^{l}$, where $l=(1),(2)$ is the layer index, and assuming the same interaction parameters for both layers, we can write,
\begin{align}
    H_{l}=&-\frac{1}{2}J^{\text{IF}}\sum_{\langle i,j \rangle}\vec{S}_i^{l}\cdot\vec{S}_j^{l}\nonumber\\&+\frac{1}{2}\sum_{\langle i,j \rangle}\vec{D}^{\text{IF}}_{ij} \cdot \qty(\vec{S}_i^{l}\times \vec{S}_j^{l})\nonumber\\&-\mu_s \sum _i \vec{B}\cdot\vec{S}_i^{l},\label{eq:layerHam}\\
    H_{\text{inter}}=&-J^{\text{IL}}\sum_{i}\vec{S}_i^{(1)}\cdot \vec{S}_i^{(2)}+\sum_{i}\vec{D}^{\text{IL}} \cdot \qty(\vec{S}_i^{(1)}\times \vec{S}_i^{(2)}).\label{eq:InteractionHam}
\end{align}
Inside each layer, we consider the nearest-neighbor interfacial exchange interaction $J^{\text{IF}}$; the nearest-neighbor interfacial DMI (IF-DMI) vector $\vec{D}^{\text{IF}}_{ij}$ perpendicular to the bonds between nearest neighbors preferring a N\'{e}el-type rotation, i.e., $\vec{D}^{\text{IF}}_{ij}= D^{\text{IF}} \vec{\hat{e}}_z \times \vec{\hat{r}}_{ij}$, where $\vec{\hat{r}}_{ij}$ is the normalized vector connecting sites $i$ and $j$; and the Zeeman interaction with an external magnetic field $\vec{B}$. $\mu_s$ is the magnetic moment at a site 
and $\langle i,j \rangle$ is the sum over all sites $i$ and their nearest neighbors $j$.
We include two interlayer couplings mediated by a non-magnetic spacer layer separating the two magnetic layers: the IEC 
$J^{\text{IL}}$ and the IL-DMI $\vec{D}^{\text{IL}}$. 
If not 
specified otherwise, we consider the parameters $J^{\text{IF}}=\SI{6}{meV}$, $|\vec{D}^{\text{IF}}|=\SI{1.5}{meV}$, and $\mu_s=3\mu_B$, where $\mu_B$ is the Bohr magneton. These values lead to the stabilization of N\'{e}el-type isolated metastable skyrmions in a collinear background inside a single layer for a sufficiently high external magnetic field applied along the out-of-plane $z$ direction, as illustrated in Fig.~\ref{fig:schematic}(a). The value of $J^{\text{IL}}$ is typically varied between $1$ and $3$~meV, while $D^{\text{IL}}$ usually takes values between $-1$ and $1$~meV. Similar ranges for the DMI parameters have been determined in Ref.~\cite{matthies2024interlayer} for Co/TM/Co(0001) heterostructures based on first-principles calculations, where TM is a single atomic layer of a transition metal Cu, Ag, Au or Pt. The interfacial exchange and DMI interactions have been demonstrated to be strongly affected by nonmagnetic capping layers, which are not explicitly included in the spin model; see, e.g., Ref.~\cite{Simon2018} for first-principles calculations on Co/Pt(111) covered by various 5$d$ elements of a single monolayer thickness, where similar $J^{\text{IF}}$ and $D^{\text{IF}}$ values have been reported to the ones used here.

To study the dynamics of the structures, we consider the 
current-perpendicular-to-the-plane (CPP) geometry. This describes the case when the multilayer is placed on a non-magnetic substrate with strong spin--orbit coupling. Besides the substrate being responsible for the interfacial DMI, an electric current flowing inside the substrate gives rise to a spin current flowing perpendicular to the layers, which exerts a spin--orbit torque on the magnetization~\cite{Manchon2019SOT}. The dynamics is described by the Landau-Lifshitz-Gilbert (LLG) equation~\cite{landau35,gilbert04} including these torque terms~\cite{aldarawsheh2024current,zhang2016magnetic},
\begin{align}
    \dot{\vec{S}}_i^{l} =& - \frac{\gamma}{1+\alpha^2} \vec{S}_i^{l} \times \left( \vec{B}^{\text{eff},l}_i + \alpha \vec{S}^{l}_i \times \vec{B}^{\text{eff},l}_i \right)\nonumber\\
    &+ \beta_f \vec{S}_i^{l} \times \left(\vec{P} + \alpha \vec{S}^{l}_i \times \vec{P} \right)\\
    &- \beta_d \vec{S}_i^{l} \times \left(\alpha \vec{P} - \vec{S}^{l}_i \times \vec{P} \right)\
    \label{eq:CPP}
\end{align}
where $\alpha$ is the Gilbert damping and $\gamma$ is the absolute value of the gyromagnetic ratio. The effective field $\vec{B}^{\text{eff},l}_i$ is given by the derivative of the Hamiltonian with respect to $\vec{S}^{l}_i$, $\vec{B}^{\text{eff},l}_i=-\mu_{s}^{-1}\dfrac{\partial H}{\partial \vec{S}^{l}_i}$; see Methods for its explicit expression. 
$\vec{P}$ is the direction of the current polarization, which will be assumed to be perpendicular to both the electric current or electric field direction inside the nonmagnetic substrate and the flow direction of the spin current perpendicular to the surface~\cite{Manchon2019SOT}, $\vec{P}\parallel\vec{j}_{\textrm{el}}\cross\hat{\vec{z}}$. The effect of the current is split into 
a field-like torque proportional to $\beta_s$ and a damping-like torque proportional to $\beta_d$. Since the field-like torque acts exactly like an external magnetic field, it distorts the skyrmions but does not set them into motion~\cite{Weienhofer2022Stochastic}. Therefore, we simplify the analysis by setting $\beta_f=0$ and only investigating the effect of the damping-like term driving the skyrmions.

\subsection{
Twin-skyrmions formed by IL-DMI}
Isolated skyrmions may be stabilized in the magnetic layers by applying an out-of-plane external magnetic field, which polarizes the spins along its direction. In the considered system, these structures are cylindrically symmetric, and may be described in the continuum limit in polar coordinates $(\rho,\phi)$ as follows,
\begin{equation}
    \Theta_{l}(\rho,\varphi)=\Theta_{l}(\rho), \quad \Phi_{l}(\rho,\varphi)=\Phi_{l}(\varphi)=m\varphi+\psi_{l},
    \label{eq:heli}
\end{equation}
where $l$ is the layer index, $\Theta$ and $\Phi$ specify the spin directions in spherical coordinates, $m$ is the winding number or vorticity of the skyrmion, and $\psi$ is its helicity. The topological charge is given as $Q=p\cdot m$, where $p$ is the polarity, i.e., the sign of the out-of-plane spin component in the center of the skyrmion. In 
Fig.~\ref{fig:schematic}, skyrmions with $m=1$ are favored by the IF-DMI, and the external field is applied along the $-z$ direction, resulting in $p=1$ and $Q=1$. 
We will primarily consider IF-DMI preferring a N\'{e}el-type rotation, resulting in $\psi=0$ or $\pi$ depending on its sign; but we will also discuss the generalization to Bloch-type rotation with $\psi=\pi/2$ or $-\pi/2$, preferred by DMI vectors along the lines connecting neighboring sites.

The possible effects of the IL-DMI on the shapes of the skyrmions are summarized in Fig.~\ref{fig:schematic}. The direction of the IL-DMI vector is highly dependent on the symmetry of the system. Both in-plane~\cite{Demiroglu2024DFTILDMI,matthies2024interlayer,Han2019} and out-of-plane~\cite{Arregi2023,pollard2020zdmi} directions have been reported, depending on the symmetry breaking introduced in the heterostructure. Here, we investigate the effect of both components. Considering only the ferromagnetic IEC, the horizontal positions of the skyrmions in the two layers become locked to each other, but their shape remains unaffected, as shown in Fig.~\ref{fig:schematic}(a). Including the IL-DMI, a finite angle is opened between the spins above each other in the two layers. For two ferromagnetic layers, this angle is given by $\alpha=\arctan \left(|\vec{D}^{\text{IL}}|/J^{\text{IL}}\right)$~\cite{matthies2024interlayer}, and the magnetic moments will align in the plane perpendicular to the IL-DMI. Here, the skyrmions in the two layers deform differently to maximize their energy gain from the IL-DMI, and form an object which we call twin-skyrmion. For example, in Fig.~\ref{fig:schematic}(b) the IL-DMI points in the $x$ direction, preferring a tilting between the two layers in the $yz$ plane. The 
background polarized along the negative $z$ direction in the top layer tilts towards the positive $y$ direction, while in the bottom layer it tilts towards the negative $y$ direction, as shown by the white arrows in Fig.~\ref{fig:schematic}(b). For a Néel-type skyrmion shown in the figure, the spins above and below the skyrmion center point along the positive or negative $y$ direction, and the system gains energy by enlarging these areas, resulting in an elongation along the $x$ direction and a small shift of the skyrmion centers in the two layers oppositely along the $y$ direction. Note that a high value of the IL-DMI $D^{\textrm{IL}}=\SI{8}{meV}$ was chosen in this figure to highlight the elongation. When the IL-DMI points along the $y$ direction instead, the polarized background tilts towards the negative and positive $x$ directions in the first and second layers, and the twin-skyrmion becomes elongated along the $y$ direction, as shown in Fig.~\ref{fig:schematic}(c). For Bloch-type skyrmions, the in-plane spin components are rotated by $90$ degrees; consequently, the elongation can also be observed perpendicular to the direction of the IL-DMI. The direction of elongation may be generally expressed as
\begin{equation}
    \hat{\vec{R}}=\pm\frac{1}{\sqrt{D_x^2+D_y^2}}\begin{pmatrix} \cos \psi & -\sin \psi \\ \sin \psi & \cos \psi \end{pmatrix} \begin{pmatrix}D^x \\ D^y\end{pmatrix},\label{eq:elongation}
\end{equation}
where $D_x$ and $D_y$ are components of the IL-DMI vector, $\psi$ is the helicity, and the $\pm$ denotes that the structure remains symmetric between the positive and negative directions. If the IL-DMI points along the $z$ direction, it prefers a rotation in the $xy$ plane, and, for a strong enough magnetic field $\vec{B}$, it does not influence the polarized background. However, the twin-skyrmion may gain energy by changing the helicities in the two layers by rotating the in-plane spin components, resulting in a negative helicity in the first and a positive helicity in the second layer in Fig.~\ref{fig:schematic}(d). The area of the skyrmion where the spins are pointing in the plane is also extended compared to Fig.~\ref{fig:schematic}(a) while retaining its circular shape.

\begin{figure}
    \centering
    \includegraphics[width=1\linewidth]{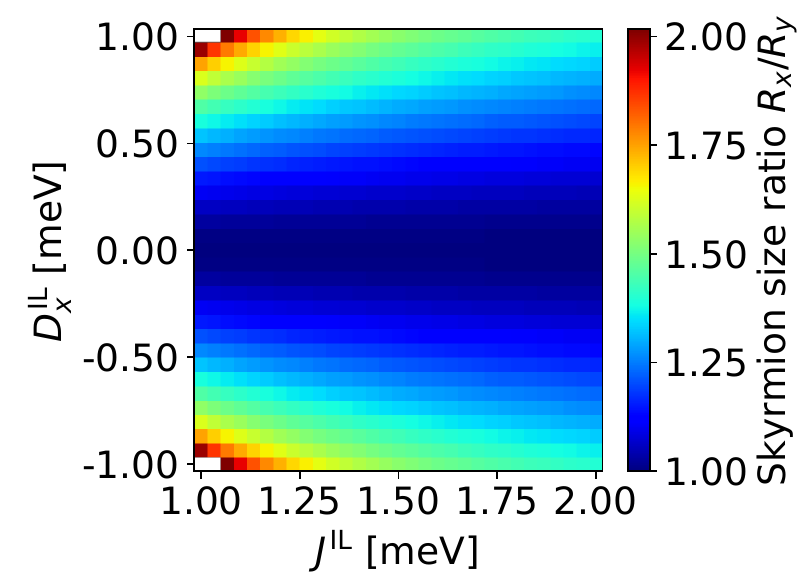}
    \caption{
    Elongation of N\'{e}el-type twin-skyrmions depending on the IEC $J^{\text{IL}}$ and IL-DMI $D_x^{\text{IL}}$. 
    The parameters are $J^{\text{IF}}=\SI{6}{meV}$, $|\vec{D}^{\text{IF}}|=\SI{1.5}{meV}$, $D_y^{\text{IL}}=D_z^{\text{IL}}=0$, 
    and $B^{z}=\SI{1.4}{T}$. For relaxing the structures, the damping parameter $\alpha=1$ was used, and the simulation was run for $5\times10^{6}$ 
    time steps, i.e., \SI{10}{ns}. 
    In the points shown in white, the elongation of the twin-skyrmion was unbounded.}
    \label{fig:skyrsize}
\end{figure}
The elongation of the twin-skyrmion for in-plane IL-DMI is quantitatively analyzed in Fig.~\ref{fig:skyrsize}. We define the twin-skyrmion radius as the distance between the skyrmion center ($\Theta(\vec{r})=\pi$) to a point where the spins lie in-plane ($\Theta(\vec{r})=\pi/2$), and take the ratio of the radii along the $x$ and $y$ directions to obtain the elongation. We only consider N\'{e}el-type skyrmions elongating along the $x$ direction for an IL-DMI along the $x$ direction; the results may be generalized to other directions of the IL-DMI based on Eq.~\eqref{eq:elongation}. Generally, $R_x/R_y$ increases with stronger IL-DMI and decreases with stronger IEC. The interactions inside the layer also hinder the elongation of the skyrmion, since this represents an energy loss compared to the circular shape preferred by these couplings. For a sufficiently strong IL-DMI, around $J^{\text{IL}}=\SI{1}{meV}$ and $D^{\text{IL}}_x=\SI{\pm 1}{meV}$ in Fig.~\ref{fig:skyrsize}, the elongation becomes unbounded. This is similar to the elliptic instability of circular isolated skyrmions~\cite{Bogdanov1999}, which may be triggered by increasing the IF-DMI. In the conventional elliptic instability, the IF-DMI prefers non-collinear structures, competing with the polarizing external field; if the IF-DMI is sufficiently strong, isolated skyrmions transform into a stripe which eventually fills up the whole area with a non-collinear structure. In contrast, the IL-DMI does not lead to structures which are modulated in the plane. However, 
spin spirals formed by the IF-DMI may gain energy from the IL-DMI by introducing a phase shift between the spirals in the two layers, while the IL-DMI competes with the out-of-plane field under all circumstances since the latter prefers to align all spins along the same direction.

\begin{figure}
    \centering
    \includegraphics[width=1\linewidth]{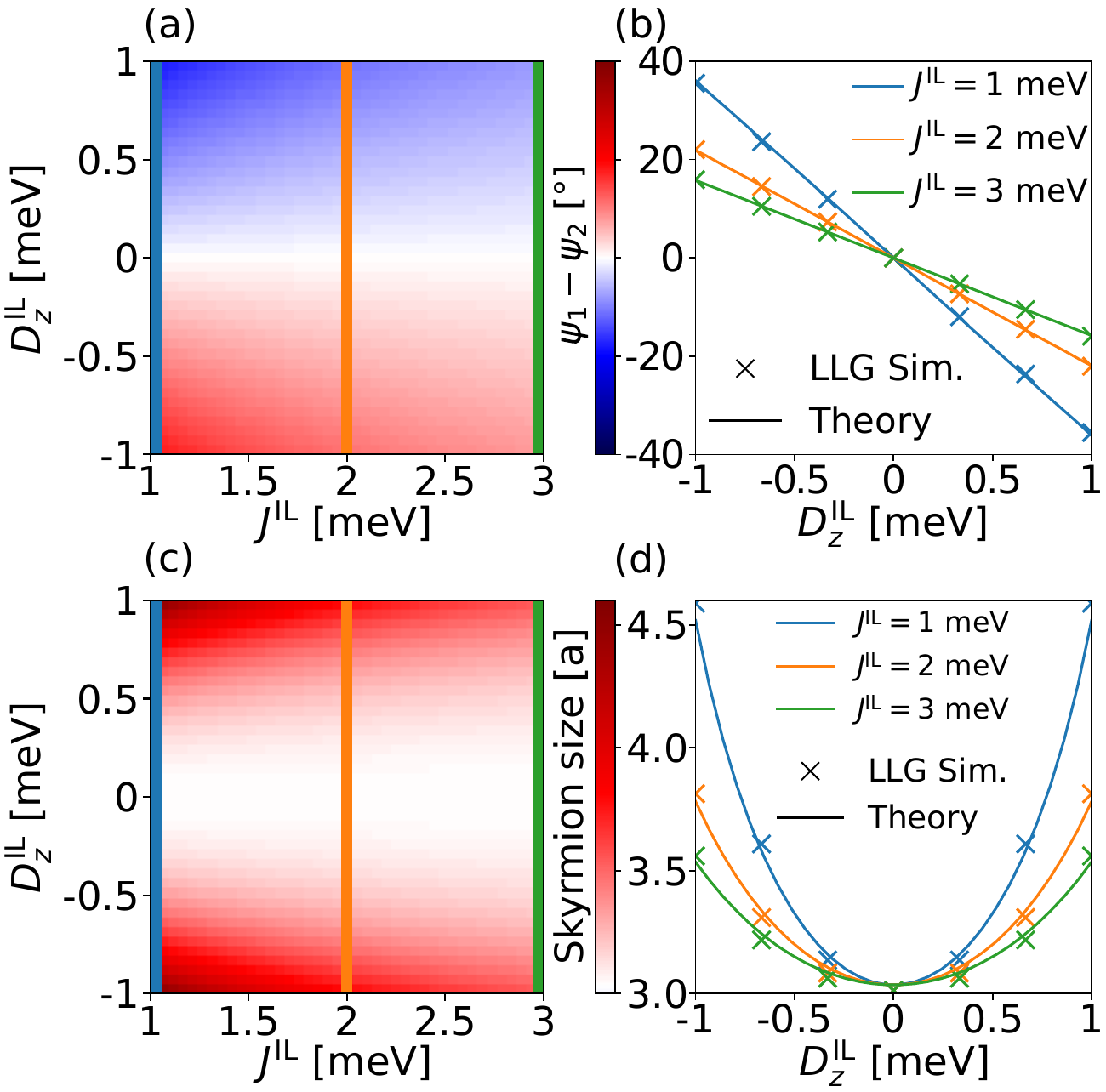}
    \caption{Change of twin-skyrmion size and helicity for out-of-plane IL-DMI. 
    (a) Difference in helicity between the layers $\psi_1-\psi_2$ as a function of $D^{\textrm{IL}}_{z}$ and $J^{\textrm{IL}}$. Panel (b) shows line cuts along the lines colored correspondingly in panel (a); crosses represent simulation results while the line shows the numerical solution of the equations in the continuum limit, see Supplementary Note~1 and Supplementary Fig.~S1. (c) Twin-skyrmion size for the same parameters, with line cuts shown in panel (d).
    The system parameters are 
    $J^{\text{IF}}=\SI{6}{meV}$, $|\vec{D}^{\text{IF}}|=\SI{1.5}{meV}$, $D_x^{\text{IL}}=D_y^{\text{IL}}=0$, and $B^{z}=\SI{3.0}{T}$. For relaxing the structures, the damping parameter $\alpha=1$ was used, and the simulation was run for $5\times10^{6}$ 
    time steps, i.e., \SI{10}{ns}. 
    }
    \label{fig:skyrangle}
\end{figure}

The influence of the out-of-plane IL-DMI on the twin-skyrmion is shown in Fig.~\ref{fig:skyrangle}. The change in the helicity from the value preferred by the IF-DMI, which is $\psi=0$ in the figure, has the same magnitude but opposite sign in the two layers. The difference between the helicities in the two layers $\psi_{1}-\psi_{2}$, i.e., twice the deviation from the $\psi=0$ value mentioned above, is illustrated in Fig.~\ref{fig:skyrangle}(a). The IL-DMI is responsible for this difference in helicity between the layers, meaning that the difference increases with $D^{\textrm{IL}}_{z}$. Changing the sign of $D^{\textrm{IL}}_{z}$ is equivalent to switching the two layers, which results in a sign change in the helicity difference. The dependence of $\psi_{1}-\psi_{2}$ on $D^{\textrm{IL}}_{z}$ is approximately linear, as shown in the line cuts Fig.~\ref{fig:skyrangle}(b). Increasing the IEC decreases the slope of the curve, and stronger interactions inside the layers also counteract this difference in helicity; e.g., the IL-DMI is competing with the IF-DMI preferring a N\'{e}el-type rotation in both layers. Since the circular shape of the skyrmion is preserved, determining the twin-skyrmion profile reduces to a radial differential equation in the continuum limit, which can be solved efficiently numerically, as presented in Supplementary Note~1. The results of this solution compare favorably to the solution of the LLG equation on the square lattice, as shown in Fig.~\ref{fig:skyrangle}(b). Increasing the IL-DMI also increases the size of the twin-skyrmion, as shown in Fig.~\ref{fig:skyrangle}(c). The size is the same for both signs of $D^{\textrm{IL}}_{z}$, and appears to depend quadratically on its magnitude, as shown in Fig.~\ref{fig:skyrangle}(d). In contrast to the in-plane IL-DMI, the ferromagnetic background does not always have a finite angle between the layers for a finite $D^{\textrm{IL}}_{z}$, since the background spins are parallel to the IL-DMI in this case. For $J^{\textrm{IL}}=\SI{1}{meV}$ and $B=\SI{3}{T}$ the IL-DMI $D^{\textrm{IL}}_{z}$ has to be larger than \SI{1.15}{meV} to overcome the magnetic field and open 
a finite angle. We restrict the values in Fig.~\ref{fig:skyrangle} to $D^{\textrm{IL}}_{z}=\pm \SI{1}{meV}$ such that the background remains parallel to the $z$ direction. We derive the formula for the threshold value of $D^{\textrm{IL}}_{z}$ in Supplementary Note~4.

\subsection{
Current-driven motion}
We study the current-driven dynamics of the twin-skyrmions in the CPP geometry. For low driving currents which do not deform the skyrmion considerably, the position of the center of mass of the skyrmions may be treated as a collective coordinate, and internal degrees of freedom may be neglected. In this approximation, the velocity $\vec{v}$ and thus the motion of the skyrmion under CPP can be described by the Thiele equation~\cite{thiele1973steady},
\begin{equation}
\vec{G}\times\vec{v}+\alpha \mathfrak{D} \cdot \vec{v}=\vec{F}=\mathcal{B}\cdot \vec{P},
\label{eq:thiele_cpp}
\end{equation}
which has been demonstrated to describe skyrmion dynamics with high accuracy~\cite{Weienhofer2022Stochastic}. Here, the gyrocoupling vector $\vec{G}$ is given by $\vec{G}=-4\pi \mu_{s}a^{-2}\gamma^{-1}Q \vec{\hat{e}}_z$, with $Q$ being the topological charge and $a$ the in-plane lattice constant. 
We take the continuum limit and consider a micromagnetic framework, $\vec{S}_i\to \vec{s}(\vec{r})$. The dissipation tensor $\mathfrak{D}$ is then given by, 
\begin{equation}
    \mathfrak{D}_{\mu \nu}= \frac{\mu_{s}}{a^{2}\gamma}\iint \partial_\mu \vec{s}(\vec{r})\cdot\partial_\nu \vec{s}(\vec{r}) \dd \vec{r},
\end{equation}
where $\mu,\nu\in \{ x,y \}$. The tensor $\mathcal{B}$ is given by,
\begin{equation}
    \mathcal{B}_{\mu k} = 
    \beta_{d}\frac{\mu_{s}}{a^{2}\gamma}\iint \left( \partial_\mu \vec{s}(\vec{r})\times \vec{s}(\vec{r}) \right)_k\dd \vec{r}.
\end{equation}
The index $\mu$ runs over the two spatial dimensions $\mu \in \{ x,y \}$ and $k$ over the three dimensions of spin direction $k \in \{x,y,z\}$. $\mathcal{B}$ transforms the three-dimensional current polarization $\vec{P}$ into the two-dimensional force $\vec{F}$ acting in the plane of the layers.

We investigate how the velocity of the twin-skyrmion is influenced by the direction and the strength of the IL-DMI. The velocity is characterized by its magnitude and direction, the latter usually expressed in terms of the skyrmion Hall angle, which is typically defined as the angle between the current driving the skyrmion and the velocity; see Supplementary Note~2. In our simulations, we fix the polarization vector $\vec{P}$ along the $x$ direction, which in the spin--orbit torque picture corresponds to an electric current $\vec{j}_{\textrm{el}}$ flowing along the positive $y$ direction. We have observed above that the IL-DMI changes the shape and size of the skyrmions, and its out-of-plane component induces a difference in the helicity between the layers. The deformation and helicity of the twin-skyrmion has no effect on the topological charge $Q$ and therefore on the gyrocoupling vector $\vec{G}$. The dissipation tensor $\mathfrak{D}$ depends on the size and shape of the skyrmions, but is independent of the helicity. The tensor $\mathcal{B}$ depends on all of the mentioned parameters. For circularly symmetric skyrmions, the latter may be expressed as a function of helicity $\psi$ and vorticity $m$ as~\cite{Weienhofer2022Stochastic},
\begin{equation}
    \mathcal{B}=f^{\text{SOT}} \begin{pmatrix}\sin \psi & -\cos \psi & 0 \\ m\cos \psi & m\sin \psi & 0\end{pmatrix}.\label{eq:Btensor}
\end{equation}

\begin{figure*}
    \centering
    \includegraphics[width=1\linewidth]{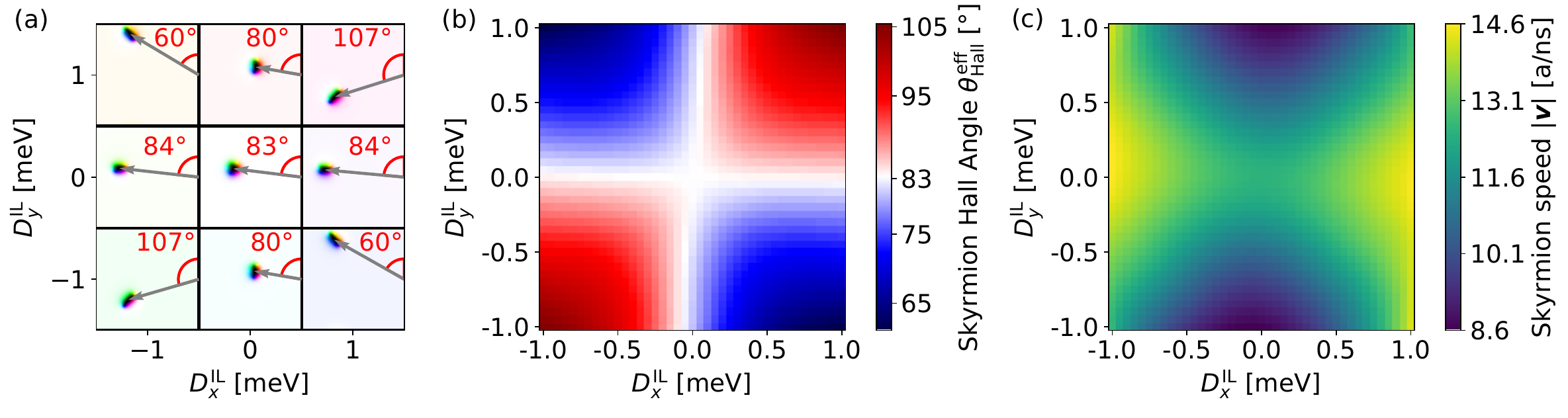}
    \caption{Velocity of twin-skyrmions for in-plane IL-DMI. 
    The current polarization $\vec{P}$ is along the $+x$ direction, corresponding to an electric current along the $+y$ direction. (a) Simulations of the current-driven motion for selected values of the IL-DMI. The arrow illustrates the displacement over the same simulation time, with the skyrmion Hall angle measured with respect to the current direction. (b) Skyrmion Hall angle and (c) magnitude of skyrmion velocity as a function of IL-DMI. The parameters are $J^{\text{IF}}=\SI{6}{meV}$, $|\vec{D}^{\text{IF}}|=\SI{1.5}{meV}$, $J^{\text{IL}}=\SI{1.75}{meV}$, $D_z^{\text{IL}}=0$, $B^{z}=\SI{1.4}{T}$, $\alpha=0.1$, and $\beta_d=\SI{1}{ns^{-1}}$. The simulation ran for $2\times10^{6}$ 
    time steps, i.e., $\SI{2}{ns}$.
    }
    \label{fig:dmixy-current}
\end{figure*}

In Fig.~\ref{fig:dmixy-current}, we study the effect of in-plane IL-DMI on N\'{e}el-type skyrmions for different directions of the IL-DMI. The current polarization along the $+x$ direction means that the results for the IL-DMI pointing along the $x$ or $y$ direction are no longer connected by rotation. Note that rotating the direction of the IL-DMI and keeping the polarization fixed is equivalent to rotating the polarization or driving current in the opposite direction for a fixed IL-DMI, which may also be performed in a material with fixed parameter values. The only symmetry observable in the figure is reversing the direction of the IL-DMI: the transformation $+\vec{D}^{\text{IL}}\to - \vec{D}^{\text{IL}}$ is the same as exchanging the top and bottom layers, $\vec{S}_i^{(1)}\to\vec{S}_i^{(2)}$ and $\vec{S}_i^{(2)}\to\vec{S}_i^{(1)}$, which does not influence the motion since the interactions inside a single layer are identical. Without IL-DMI, the skyrmion Hall angle measured from the nominal current direction along $+y$ is around $83^{\circ}$, while the velocity is $13a\cdot\textrm{ns}^{-1}$. If the IL-DMI is approximately parallel to the $x$ or $y$ axis, the skyrmion Hall angle only minimally varies, but the velocities may differ by up to 70\% for the same driving current and magnitude of IL-DMI. In particular, the skyrmions move faster along the direction they are elongated due to the IL-DMI, and slower if the direction of movement determined by the current polarization is approximately perpendicular to the elongation. Along the lines $D_x^{\text{IL}}=\pm D_y^{\text{IL}}$, a different effect can be observed: the velocity is approximately constant, but the skyrmion Hall angle differs considerably, taking the value of $60^{\circ}$ for $D_x^{\text{IL}}=- D_y^{\text{IL}}=\pm\SI{1}{meV}$ and $107^{\circ}$ for $D_x^{\text{IL}}=D_y^{\text{IL}}=\pm\SI{1}{meV}$. The angle between the movement and the elongation directions is similarly low along both of these lines, resulting in a relatively high velocity. Overall, it can be observed that for low angles between the direction of elongation and the current polarization, the Hall angle of the twin-skyrmion is adjusted to roughly follow the elongation direction while keeping a high velocity, whereas for higher angles between these two directions the movement slows down and the skyrmion Hall angle returns to its value for uncoupled layers. This enables tuning the skyrmion Hall angle and the velocity separately from each other.

\begin{figure}
    \centering
    \includegraphics[width=1\linewidth]{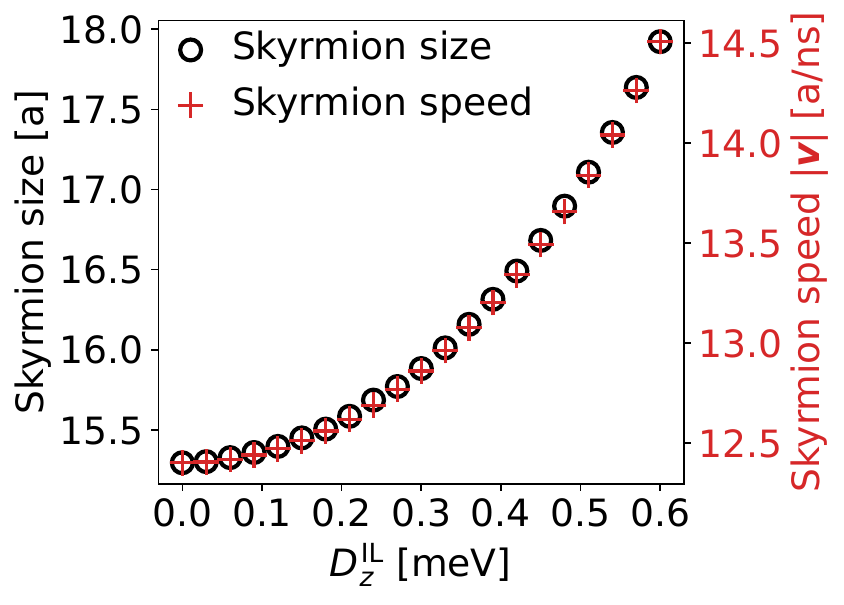}
    \caption{Effect of the $D^{\text{IL}}_z$ on the twin-skyrmion velocity and size. 
    Values of $D^{\text{IL}}_z$ greater than $\SI{0.6}{meV}$ do not result in stable motion. The system parameters are: $J^{\text{IL}}=\SI{1.75}{meV}$, $D^{\text{IL}}_x=D^{\text{IL}}_y=0$, $\vec{B}=\SI{1.4}{T}$, $\alpha=0.1$, and $\beta_d=\SI{1}{ns^{-1}}$. The simulation ran for $2\times10^{6}$ time steps, i.e., $\SI{2}{ns}$.}
    \label{fig:dmiz_speed_angle}
\end{figure}

The influence of the out-of-plane component of the IL-DMI is investigated in Fig.~\ref{fig:dmiz_speed_angle}. As it was shown in Fig.~\ref{fig:skyrangle}, this type of IL-DMI causes a difference in the helicity between the skyrmions in the two layers. This would induce different velocity directions for the two parts of the twin-skyrmion, which is expected to cause an instability. This can also be observed in the simulations, because only for small IL-DMI values ($\vec{D}^{\text{IL}}_z<\SI{0.6}{meV}$ for the chosen parameters) is the motion of twin-skyrmions stable. It could be expected that the velocity of the twin-skyrmion is determined by the average of the velocity vectors in the two layers, thus the different velocity directions between the layers leads to a reduction of the net velocity. On the contrary, an increase in the velocity with $D_{z}^{\textrm{IL}}$ may be observed in Fig.~\ref{fig:dmiz_speed_angle}. This is caused by an increase in the twin-skyrmion size with IL-DMI, as shown in Figs.~\ref{fig:skyrangle} and \ref{fig:dmiz_speed_angle}. The force calculated from the $\mathcal{B}$ tensor increases approximately linearly with skyrmion size; see Supplementary Note~3 for the derivation and Supplementary Fig.~S2. Since the changes in the dissipation tensor $\mathfrak{D}$ are small compared to $\mathcal{B}$, the velocity of the twin-skyrmion correlates very closely with the skyrmion size and the force acting on the skyrmion. Since the dissipation tensor depends weakly on the IL-DMI in this case, the Hall angle only increases by 0.3\% when increasing $D^{\text{IL}}_z$ from \SI{0.0}{meV} to \SI{0.6}{meV}.

\section{\label{sec:conclusion}Discussion}

In this paper, we studied magnetic skyrmions in bilayer systems coupled by interlayer Heisenberg and Dzyaloshinskii--Moriya interactions. We found that the IL-DMI locks the skyrmions in the two layers to each other but also deforms their shape, thereby stabilizing a structure which we termed twin-skyrmion. The in-plane IL-DMI tilts the magnetization in the collinear regions in the two layers in opposite directions. It also causes the skyrmions to elongate and their centers to shift oppositely in the two layers, thereby maximizing the energy gain from this energy term. For a sufficiently high value of the IL-DMI, this results in an elliptic instability of the twin-skyrmion. The out-of-plane IL-DMI preserves the circular shape of the skyrmions in the two layers, but increases their radius and changes their helicities in opposite directions.

Furthermore, we studied how the IL-DMI influences the current-driven motion of twin-skyrmions in the CPP geometry. For in-plane IL-DMI, we found that the dynamics strongly depends on the relative directions of the equilibrium elongation and the current polarization. The twin-skyrmion prefers to move along the direction of its elongation by increasing its velocity compared to uncoupled skyrmions in the two layers and adjusting its skyrmion Hall angle if the polarization direction is changed. However, if the direction of motion preferred by the current polarization is approximately perpendicular to the elongation, the velocity reduces and the skyrmion Hall angle stays close to its value for uncoupled layers. For out-of-plane IL-DMI, we observed an increase in skyrmion velocity together with the skyrmion size, while the skyrmion Hall angle was found to be hardly affected.

These findings could motivate experimental studies on skyrmions in magnetic multilayers coupled by IL-DMI, expanding upon previous works in systems coupled by ferromagnetic or antiferromagnetic IEC. 
The possibility to change the skyrmion velocity or the Hall angle separately depending on the current polarization direction should provide improved control over the current-driven motion of skyrmions. It has been demonstrated that skyrmions with circular equilibrium profiles become distorted by even stronger driving currents~\cite{masell2020SkyrDeformation,Liu2020DeformedSkyrm,lee2025skyrmionhalleffectshape}. The dynamics in this strongly nonlinear regime is expected to become even more complex in the presence of IL-DMI, which may prefer a different direction of elongation compared to the driving current and also affects the stability of twin-skyrmions.

\section{\label{sec:theory}Methods}

The effective field $\vec{B}^{\text{eff},l}_i$ in the LLG equation~\eqref{eq:CPP} is given in the two layers as
\begin{align}
    \vec{B}^{\text{eff},l}_i=&\frac{J^{\text{IF}}}{\mu_s}\sum_{\langle j \rangle_i} \vec{S}_j^{l}+\frac{1}{\mu_s}\sum _{\langle j \rangle_i} \vec{D}_{ij}^{\text{IF}} \times \vec{S}_j^{l}+\vec{B}\nonumber\\
    &+\frac{J^{\text{IL}}}{\mu_s} \vec{S}_i^{\overline{l}}\pm\frac{1}{\mu_s} \vec{D}^{\text{IL}} \times \vec{S}^{\overline{l}}_i,
\end{align}
where $\langle j \rangle_i$ denotes the sum over all nearest neighbors $j$ of the site $i$, and $\overline{l}$ denotes the layer other than $l$. 
The positive sign is taken in the last term for $l=(1)$ and the negative sign for $l=(2)$ because of the definition of the cross product.

The simulation is performed on a $128\times 128$ square lattice with periodic boundary conditions for each layer. The used integration scheme is the Euler method with an additional normalization step after each iteration. The time step is $\Delta t=\SI{1e-6}{ns}$ for Figs.~\ref{fig:schematic}, 
\ref{fig:dmixy-current}, and 
\ref{fig:dmiz_speed_angle}. Some simulations were repeated with various time steps to confirm the stability of the solution for the selected time step. For the static properties in Figs.~\ref{fig:skyrsize} and 
\ref{fig:skyrangle}, a larger time step $\Delta t=\SI{2e-6}{ns}$ was used, since here the focus is on obtaining a local energy minimum instead of following the time evolution. The simulation was always terminated after a fixed number of iterations. The number of iterations is given in each figure caption. Close to the boundary of the stability region in Fig.~\ref{fig:skyrsize}, we confirmed the stability by increasing the simulation time by a factor of ten, and we did not observe a measurable change in the skyrmion size, with this size staying well below the size of the simulation cell. Outside the stability region, we increased the system size by a factor of two, and observed a further elongation of the skyrmion until it filled up the simulation cell and was stopped by confinement.

\section*{Data availability}
The datasets generated or analysed during the current study are not publicly available due to it being not technically feasible or the cost of preparing, depositing, and hosting the data would be prohibitive within the terms of this research project, but are available from the corresponding author on reasonable request.

\section*{Code availability}
The underlying code for this study is not publicly available but may be made available to qualified researchers on reasonable request from the corresponding author.

\bibliography{references}

@article{zhang2016magnetic,
  title={Magnetic bilayer-skyrmions without skyrmion {H}all effect},
  author={Zhang, Xichao and Zhou, Yan and Ezawa, Motohiko},
  journal={Nat. Commun.},
  volume={7},
  number={1},
  pages={10293},
  year={2016},
  publisher={Nature Publishing Group UK London},
  doi={10.1038/ncomms10293},
  url={https://doi.org/10.1038/ncomms10293}
}

@article{aldarawsheh2024current,
  title={{Current-driven dynamics of antiferromagnetic skyrmions: from skyrmion Hall effects to hybrid inter-skyrmion scattering}},
  author={Aldarawsheh, Amal and Sallermann, Moritz and Abusaa, Muayad and Lounis, Samir},
  journal={npj Spintronics},
  volume={2},
  number={1},
  pages={41},
  year={2024},
  publisher={Nature Publishing Group UK London},
  doi = {10.1038/s44306-024-00049-w},
  url = {https://doi.org/10.1038/s44306-024-00049-w},
}

@article{xia2020elliptical,
    author = {Xia, Jing and others},
    title = {Dynamics of an elliptical ferromagnetic skyrmion driven by the spin–orbit torque},
    journal = {Applied Physics Letters},
    volume = {116},
    number = {2},
    pages = {022407},
    year = {2020},
    month = {01},
    issn = {0003-6951},
    doi = {10.1063/1.5132915},
    url = {https://doi.org/10.1063/1.5132915},
}

@article{matthies2024interlayer,
  title = {Interlayer and interfacial {D}zyaloshinskii-{M}oriya interaction in magnetic trilayers: First-principles calculations},
  author = {Matthies, Tim and R\'ozsa, Levente and Szunyogh, L\'aszl\'o and Wiesendanger, Roland and Vedmedenko, Elena Y.},
  journal = {Phys. Rev. Res.},
  volume = {6},
  issue = {4},
  pages = {043158},
  numpages = {16},
  year = {2024},
  month = {Nov},
  publisher = {American Physical Society},
  doi = {10.1103/PhysRevResearch.6.043158},
  url = {https://link.aps.org/doi/10.1103/PhysRevResearch.6.043158}
}

@article{thiele1973steady,
  title = {Steady-State Motion of Magnetic Domains},
  author = {Thiele, A. A.},
  journal = {Phys. Rev. Lett.},
  volume = {30},
  issue = {6},
  pages = {230--233},
  numpages = {0},
  year = {1973},
  month = {Feb},
  publisher = {American Physical Society},
  doi = {10.1103/PhysRevLett.30.230},
  url = {https://link.aps.org/doi/10.1103/PhysRevLett.30.230}
}

@phdthesis{Weienhofer2022Stochastic,
  year={2022},
  title={Stochastic and Deterministic Dynamics of Topological Spin Textures : Theory and Simulation},
  author={Weißenhofer, Markus},
  address={Konstanz},
  school={Universität Konstanz}
}

@article{jiang2017direct,
  title = {{Direct observation of the skyrmion Hall effect}},
  volume = {13},
  ISSN = {1745-2481},
  url = {http://dx.doi.org/10.1038/nphys3883},
  DOI = {10.1038/nphys3883},
  number = {2},
  journal = {Nature Physics},
  publisher = {Springer Science and Business Media LLC},
  author = {Jiang,  Wanjun and others},
  year = {2016},
  month = sep,
  pages = {162–169}
}

@ARTICLE{2020SciPy-NMeth,
  author  = {Virtanen, Pauli and others},
  title   = {{{SciPy} 1.0: Fundamental Algorithms for Scientific
            Computing in Python}},
  journal = {Nature Methods},
  year    = {2020},
  volume  = {17},
  pages   = {261--272},
  adsurl  = {https://rdcu.be/b08Wh},
  doi     = {10.1038/s41592-019-0686-2},
}

@article{bvp2001,
author = {Kierzenka, Jacek and Shampine, Lawrence F.},
title = {A {BVP} solver based on residual control and the Maltab {PSE}},
year = {2001},
issue_date = {September 2001},
publisher = {Association for Computing Machinery},
address = {New York, NY, USA},
volume = {27},
number = {3},
issn = {0098-3500},
url = {https://doi.org/10.1145/502800.502801},
doi = {10.1145/502800.502801},
journal = {ACM Trans. Math. Softw.},
month = sep,
pages = {299–316},
numpages = {18}
}

@article{parkin2021,
author = {Yang, Hun and Naaman, Ron and Paltiel, Yossi and Parkin, Stuart P.},
title = {Chiral spintronics},
year = {2021},
publisher = {Nature London},
address = {New York, NY, USA},
volume = {3},
number = {3},
issn = {0098-3500},
doi = {10.1038/s42254-021-00302-9},
journal = {Nat. Rev. Phys.},
month = may,
pages = {328-343}
}

@article{pai2024,
author = {Lin, Chun-Yi and others},
title = {Field-Free Spin–Orbit Torque Switching via Oscillatory Interlayer {D}zyaloshinskii–{M}oriya Interaction for Advanced Memory Applications},
journal = {ACS Mater. Lett.},
volume = {6},
number = {2},
pages = {400-408},
year = {2024},
doi = {10.1021/acsmaterialslett.3c01376},
}

@article{psaroudaki2022,
author = {Psaroudaki, C. and Panagopoulos,  C. },
title = {Skyrmion helicity: Quantization and quantum tunneling effects},
year = {2022},
publisher = {APS},
address = {New York, NY, USA},
volume = {106},
doi = {10.1103/PhysRevB.106.104422},
journal = {Phys. Rev. B},
pages = {104422}
}

@article{liu2024,
author = {Liu, Y. and Yang,  B. and Guo X. and Picozzi, S. and Yan, Y. },
title = {{Modulation of skyrmion helicity by competition between Dzyaloshinskii-Moriya
interaction and magnetic frustration}},
year = {2024},
publisher = {APS},
address = {New York, NY, USA},
volume = {109},
doi = {10.1103/PhysRevB.109.094431},
journal = {Phys. Rev. B},
pages = {094431}
}

@article{kammerbauer2023,
author = {Kammerbauer, F. and others },
title = {{Controlling the interlayer Dzyaloshinskii-Moriya by electrical currents}},
year = {2023},
publisher = {ACS},
address = {New York, NY, USA},
volume = {23},
doi = {},
journal = {Nano Lett.},
pages = {7070-7075},
doi = {10.1021/acs.nanolett.3c01709},
URL = { https://doi.org/10.1021/acs.nanolett.3c01709},
}

@article{Manchon2019SOT,
  title = {Current-induced spin-orbit torques in ferromagnetic and antiferromagnetic systems},
  author = {Manchon, A. and others},
  journal = {Rev. Mod. Phys.},
  volume = {91},
  issue = {3},
  pages = {035004},
  numpages = {80},
  year = {2019},
  month = {Sep},
  publisher = {American Physical Society},
  doi = {10.1103/RevModPhys.91.035004},
  url = {https://link.aps.org/doi/10.1103/RevModPhys.91.035004}
}

@article{Qiu2021SynthAntiSkyr,
    author = {Qiu, Lei and others},
    title = {Interlayer coupling effect on skyrmion dynamics in synthetic antiferromagnets},
    journal = {Appl. Phys. Lett.},
    volume = {118},
    number = {8},
    pages = {082403},
    year = {2021},
    month = {02},
    issn = {0003-6951},
    doi = {10.1063/5.0039470},
    url = {https://doi.org/10.1063/5.0039470},
}

@article{Matheus2024SkyrAntiferro,
  title = {Stability and dynamics of synthetic antiferromagnetic skyrmions in asymmetric multilayers},
  author = {Correia, Matheus V. and Vel\'asquez, J. C. P. and de Souza Silva, Cl\'ecio C.},
  journal = {Phys. Rev. B},
  volume = {110},
  issue = {9},
  pages = {094430},
  numpages = {9},
  year = {2024},
  month = {Sep},
  publisher = {American Physical Society},
  doi = {10.1103/PhysRevB.110.094430},
  url = {https://link.aps.org/doi/10.1103/PhysRevB.110.094430}
}

@article{juge2022skyrmions,
  title={Skyrmions in synthetic antiferromagnets and their nucleation via electrical current and ultra-fast laser illumination},
  author={Juge, Rom{\'e}o and others},
  journal={Nat. Commun},
  volume={13},
  number={1},
  pages={4807},
  year={2022},
  publisher={Nature Publishing Group UK London},
  doi={10.1038/s41467-022-32525-4},
  url={https://doi.org/10.1038/s41467-022-32525-4}
}

@article{legrand2020room,
  title={Room-temperature stabilization of antiferromagnetic skyrmions in synthetic antiferromagnets},
  author={Legrand, William and others},
  journal={Nat. Mater.},
  volume={19},
  number={1},
  pages={34--42},
  year={2020},
  publisher={Nature Publishing Group UK London},
  doi={10.1038/s41563-019-0468-3},
  url={https://doi.org/10.1038/s41563-019-0468-3}
}

@article{masell2020SkyrDeformation,
  title = {Spin-transfer torque driven motion, deformation, and instabilities of magnetic skyrmions at high currents},
  author = {Masell, J. and Rodrigues, D. R. and McKeever, B. F. and Everschor-Sitte, K.},
  journal = {Phys. Rev. B},
  volume = {101},
  issue = {21},
  pages = {214428},
  numpages = {13},
  year = {2020},
  month = {Jun},
  publisher = {American Physical Society},
  doi = {10.1103/PhysRevB.101.214428},
  url = {https://link.aps.org/doi/10.1103/PhysRevB.101.214428}
}

@article{Liu2020DeformedSkyrm,
  title = {Current-Driven Skyrmion Motion Beyond Linear Regime: Interplay between Skyrmion Transport and Deformation},
  author = {Liu, Linjie and Chen, Weijin and Zheng, Yue},
  journal = {Phys. Rev. Appl.},
  volume = {14},
  issue = {2},
  pages = {024077},
  numpages = {15},
  year = {2020},
  month = {Aug},
  publisher = {American Physical Society},
  doi = {10.1103/PhysRevApplied.14.024077},
  url = {https://link.aps.org/doi/10.1103/PhysRevApplied.14.024077}
}

@misc{lee2025skyrmionhalleffectshape,
      title={{Skyrmion Hall effect and shape deformation of current-driven bilayer skyrmions in synthetic antiferromagnets}}, 
      author={Mu-Kun Lee and Javier A. Vélez and Rubén M. Otxoa and Masahito Mochizuki},
      year={2025},
      eprint={2507.15531},
      archivePrefix={arXiv},
      primaryClass={cond-mat.mes-hall},
      url={https://arxiv.org/abs/2507.15531}, 
}

@article{xia2021TubesSynthAntiFM,
  title = {Current-induced dynamics of skyrmion tubes in synthetic antiferromagnetic multilayers},
  author = {Xia, Jing and others},
  journal = {Phys. Rev. B},
  volume = {103},
  issue = {17},
  pages = {174408},
  numpages = {8},
  year = {2021},
  month = {May},
  publisher = {American Physical Society},
  doi = {10.1103/PhysRevB.103.174408},
  url = {https://link.aps.org/doi/10.1103/PhysRevB.103.174408}
}

@article{zhao2025interlayer,
  title={{Interlayer Dzyaloshinskii-Moriya interaction in spintronics}},
  author={Zhao, Xupeng and Zhao, Jianhua},
  journal={npj Spintronics},
  volume={3},
  number={1},
  pages={25},
  year={2025},
  publisher={Nature Publishing Group UK London},
  doi={10.1038/s44306-025-00091-2}
}

@article{fernandez2017three,
  title={Three-dimensional nanomagnetism},
  author={Fern{\'a}ndez-Pacheco, Amalio and others},
  journal={Nat. Commun.},
  volume={8},
  number={1},
  pages={15756},
  year={2017},
  publisher={Nature Publishing Group UK London},
  doi={10.1038/ncomms15756}
}

@article{Grollier2020,
  title = {Neuromorphic spintronics},
  volume = {3},
  ISSN = {2520-1131},
  url = {http://dx.doi.org/10.1038/s41928-019-0360-9},
  DOI = {10.1038/s41928-019-0360-9},
  number = {7},
  journal = {Nat. Electron.},
  publisher = {Springer Science and Business Media LLC},
  author = {Grollier,  J. and others},
  year = {2020},
  month = mar,
  pages = {360–370}
}

@article{Goebel2021,
  title = {{Beyond skyrmions: Review and perspectives of alternative magnetic quasiparticles}},
  volume = {895},
  ISSN = {0370-1573},
  url = {http://dx.doi.org/10.1016/j.physrep.2020.10.001},
  DOI = {10.1016/j.physrep.2020.10.001},
  journal = {Phys. Rep.},
  publisher = {Elsevier BV},
  author = {G\"{o}bel,  B\"{o}rge and Mertig,  Ingrid and Tretiakov,  Oleg A.},
  year = {2021},
  month = feb,
  pages = {1–28}
}

@article{Marrows2024,
  title = {Neuromorphic computing with spintronics},
  volume = {2},
  pages = {12},
  ISSN = {2948-2119},
  url = {http://dx.doi.org/10.1038/s44306-024-00019-2},
  DOI = {10.1038/s44306-024-00019-2},
  number = {1},
  journal = {npj Spintronics},
  publisher = {Springer Science and Business Media LLC},
  author = {Marrows,  Christopher H. and Barker,  Joseph and Moore,  Thomas A. and Moorsom,  Timothy},
  year = {2024},
  month = apr 
}

@article{Makarov2021,
  title = {{New Dimension in Magnetism and Superconductivity: 3D and Curvilinear Nanoarchitectures}},
  volume = {34},
  pages = {202101758},
  ISSN = {1521-4095},
  url = {http://dx.doi.org/10.1002/adma.202101758},
  DOI = {10.1002/adma.202101758},
  number = {3},
  journal = {Adv. Mater.},
  publisher = {Wiley},
  author = {Makarov,  Denys and others},
  year = {2021},
  month = oct 
}

@article{Vedmedenko2022,
  title = {{Interlayer Dzyaloshinskii-Moriya Interactions}},
  author = {Vedmedenko, Elena Y. and Riego, Patricia and Arregi, Jon Ander and Berger, Andreas},
  journal = {Phys. Rev. Lett.},
  volume = {122},
  issue = {25},
  pages = {257202},
  numpages = {5},
  year = {2019},
  month = {Jun},
  publisher = {American Physical Society},
  doi = {10.1103/PhysRevLett.122.257202},
  url = {https://link.aps.org/doi/10.1103/PhysRevLett.122.257202}
}

@article{Han2019,
  title = {Long-range chiral exchange interaction in synthetic antiferromagnets},
  volume = {18},
  ISSN = {1476-4660},
  url = {http://dx.doi.org/10.1038/s41563-019-0370-z},
  DOI = {10.1038/s41563-019-0370-z},
  number = {7},
  journal = {Nat. Mater.},
  publisher = {Springer Science and Business Media LLC},
  author = {Han,  Dong-Soo and others},
  year = {2019},
  month = jun,
  pages = {703–708}
}

@article{Arregi2023,
  title = {{Large interlayer Dzyaloshinskii-Moriya interactions across Ag-layers}},
  volume = {14},
  pages = {6927},
  ISSN = {2041-1723},
  url = {http://dx.doi.org/10.1038/s41467-023-42426-9},
  DOI = {10.1038/s41467-023-42426-9},
  number = {1},
  journal = {Nat. Commun.},
  publisher = {Springer Science and Business Media LLC},
  author = {Arregi,  Jon Ander and Riego,  Patricia and Berger,  Andreas and Vedmedenko,  Elena Y.},
  year = {2023},
  month = oct 
}

@article{CascalesSandoval2025,
  title = {Preservation of scalar spin chirality across a metallic spacer in synthetic antiferromagnets with chiral interlayer interactions},
  author = {Cascales Sandoval, Miguel A. and others},
  journal = {Phys. Rev. B},
  volume = {112},
  issue = {2},
  pages = {024413},
  numpages = {9},
  year = {2025},
  month = {Jul},
  publisher = {American Physical Society},
  doi = {10.1103/9798-rdmj},
  url = {https://link.aps.org/doi/10.1103/9798-rdmj}
}

@article{Dzyaloshinsky1958,
  title = {A thermodynamic theory of “weak” ferromagnetism of antiferromagnetics},
  volume = {4},
  ISSN = {0022-3697},
  url = {http://dx.doi.org/10.1016/0022-3697(58)90076-3},
  DOI = {10.1016/0022-3697(58)90076-3},
  number = {4},
  journal = {J. Phys. Chem. Solids},
  publisher = {Elsevier BV},
  author = {Dzyaloshinsky,  I.},
  year = {1958},
  month = jan,
  pages = {241–255}
}

@article{Moriya1960,
  title = {{Anisotropic Superexchange Interaction and Weak Ferromagnetism}},
  author = {Moriya, T\^oru},
  journal = {Phys. Rev.},
  volume = {120},
  issue = {1},
  pages = {91--98},
  numpages = {0},
  year = {1960},
  month = {Oct},
  publisher = {American Physical Society},
  doi = {10.1103/PhysRev.120.91},
  url = {https://link.aps.org/doi/10.1103/PhysRev.120.91}
}

@article{Legrand2018,
  title = {Hybrid chiral domain walls and skyrmions in magnetic multilayers},
  volume = {4},
  pages = {aat0415},
  ISSN = {2375-2548},
  url = {http://dx.doi.org/10.1126/sciadv.aat0415},
  DOI = {10.1126/sciadv.aat0415},
  number = {7},
  journal = {Sci. Adv.},
  publisher = {American Association for the Advancement of Science (AAAS)},
  author = {Legrand,  William and others},
  year = {2018},
  month = jul 
}

@article{Dohi2019,
  title = {Formation and current-induced motion of synthetic antiferromagnetic skyrmion bubbles},
  volume = {10},
  pages = {5153},
  ISSN = {2041-1723},
  url = {http://dx.doi.org/10.1038/s41467-019-13182-6},
  DOI = {10.1038/s41467-019-13182-6},
  number = {1},
  journal = {Nat. Commun.},
  publisher = {Springer Science and Business Media LLC},
  author = {Dohi,  Takaaki and DuttaGupta,  Samik and Fukami,  Shunsuke and Ohno,  Hideo},
  year = {2019},
  month = nov 
}

@article{Dohi2023,
  title = {Enhanced thermally-activated skyrmion diffusion with tunable effective gyrotropic force},
  volume = {14},
  pages = {5424},
  ISSN = {2041-1723},
  url = {http://dx.doi.org/10.1038/s41467-023-40720-0},
  DOI = {10.1038/s41467-023-40720-0},
  number = {1},
  journal = {Nat. Commun.},
  publisher = {Springer Science and Business Media LLC},
  author = {Dohi,  Takaaki and others},
  year = {2023},
  month = sep 
}

@article{Guo2022,
  title = {{Effect of interlayer Dzyaloshinskii-Moriya interaction on spin structure in synthetic antiferromagnetic multilayers}},
  author = {Guo, Yaqin and others},
  journal = {Phys. Rev. B},
  volume = {105},
  issue = {18},
  pages = {184405},
  numpages = {9},
  year = {2022},
  month = {May},
  publisher = {American Physical Society},
  doi = {10.1103/PhysRevB.105.184405},
  url = {https://link.aps.org/doi/10.1103/PhysRevB.105.184405}
}

@article{Back2020,
  title = {The 2020 skyrmionics roadmap},
  volume = {53},
  ISSN = {1361-6463},
  url = {http://dx.doi.org/10.1088/1361-6463/ab8418},
  DOI = {10.1088/1361-6463/ab8418},
  number = {36},
  journal = {J. Phys. D},
  publisher = {IOP Publishing},
  author = {Back,  C and others},
  year = {2020},
  month = jun,
  pages = {363001}
}

@article{Azhar2022,
  title = {{Screw Dislocations in Chiral Magnets}},
  author = {Azhar, Maria and Kravchuk, Volodymyr P. and Garst, Markus},
  journal = {Phys. Rev. Lett.},
  volume = {128},
  issue = {15},
  pages = {157204},
  numpages = {6},
  year = {2022},
  month = {Apr},
  publisher = {American Physical Society},
  doi = {10.1103/PhysRevLett.128.157204},
  url = {https://link.aps.org/doi/10.1103/PhysRevLett.128.157204}
}

@article{Zheng2023,
  title = {Hopfion rings in a cubic chiral magnet},
  volume = {623},
  ISSN = {1476-4687},
  url = {http://dx.doi.org/10.1038/s41586-023-06658-5},
  DOI = {10.1038/s41586-023-06658-5},
  number = {7988},
  journal = {Nature},
  publisher = {Springer Science and Business Media LLC},
  author = {Zheng,  Fengshan and others},
  year = {2023},
  month = nov,
  pages = {718–723}
}

@article{Nagaosa2013,
  title = {Topological properties and dynamics of magnetic skyrmions},
  volume = {8},
  ISSN = {1748-3395},
  url = {http://dx.doi.org/10.1038/nnano.2013.243},
  DOI = {10.1038/nnano.2013.243},
  number = {12},
  journal = {Nat. Nanotechnol.},
  publisher = {Springer Science and Business Media LLC},
  author = {Nagaosa,  Naoto and Tokura,  Yoshinori},
  year = {2013},
  month = dec,
  pages = {899–911}
}

@article{Msiska2022,
  title = {{Nonzero Skyrmion Hall Effect in Topologically Trivial Structures}},
  author = {Msiska, Robin and Rodrigues, Davi R. and Leliaert, Jonathan and Everschor-Sitte, Karin},
  journal = {Phys. Rev. Appl.},
  volume = {17},
  issue = {6},
  pages = {064015},
  numpages = {6},
  year = {2022},
  month = {Jun},
  publisher = {American Physical Society},
  doi = {10.1103/PhysRevApplied.17.064015},
  url = {https://link.aps.org/doi/10.1103/PhysRevApplied.17.064015}
}

@article{Hrabec2017,
  title = {Current-induced skyrmion generation and dynamics in symmetric bilayers},
  volume = {8},
  pages = {15765},
  ISSN = {2041-1723},
  url = {http://dx.doi.org/10.1038/ncomms15765},
  DOI = {10.1038/ncomms15765},
  number = {1},
  journal = {Nat. Commun.},
  publisher = {Springer Science and Business Media LLC},
  author = {Hrabec,  A. and others},
  year = {2017},
  month = jun 
}

@article{landau35,
  author =        {Landau, L D and Lifshitz, E M},
  journal =       {Phys. Z. Sowjetunion},
  pages =         {153--169},
  title =         {On the theory of the dispersion of magnetic
                   permeability in ferromagnetic bodies},
  volume =        {8},
  year =          {1935},
  doi = {https://doi.org/10.1016/B978-0-08-036364-6.50008-9},
  url = {https://www.sciencedirect.com/science/article/pii/B9780080363646500089},
}

@article{gilbert04,
  author =        {Gilbert, T. L.},
  journal =       {IEEE Trans. Magn.},
  month =         {Nov},
  number =        {6},
  pages =         {3443-3449},
  title =         {A phenomenological theory of damping in ferromagnetic
                   materials},
  volume =        {40},
  year =          {2004},
  doi =           {10.1109/TMAG.2004.836740},
  issn =          {0018-9464},
}

@article{Bogdanov1999,
  title = {The stability of vortex-like structures in uniaxial ferromagnets},
  volume = {195},
  ISSN = {0304-8853},
  url = {http://dx.doi.org/10.1016/S0304-8853(98)01038-5},
  DOI = {10.1016/s0304-8853(98)01038-5},
  number = {1},
  journal = {Journal of Magnetism and Magnetic Materials},
  publisher = {Elsevier BV},
  author = {Bogdanov,  A. and Hubert,  A.},
  year = {1999},
  month = apr,
  pages = {182–192}
}

@article{rozsa2017TempScaling,
  title = {Temperature scaling of the {Dzyaloshinsky-Moriya} interaction in the spin wave spectrum},
  author = {R\'ozsa, Levente and Atxitia, Unai and Nowak, Ulrich},
  journal = {Phys. Rev. B},
  volume = {96},
  issue = {9},
  pages = {094436},
  numpages = {11},
  year = {2017},
  month = {Sep},
  publisher = {American Physical Society},
  doi = {10.1103/PhysRevB.96.094436},
  url = {https://link.aps.org/doi/10.1103/PhysRevB.96.094436}
}

@article{DEGER2019165399,
title = {Impact of interlayer coupling on magnetic skyrmion size},
journal = {J. of Magn. Magn. Mater.},
volume = {489},
pages = {165399},
year = {2019},
issn = {0304-8853},
doi = {https://doi.org/10.1016/j.jmmm.2019.165399},
url = {https://www.sciencedirect.com/science/article/pii/S0304885318326763},
author = {C. Deger and I. Yavuz and F. Yildiz},
}

@article{Schrautzer2022,
  title = {Effects of interlayer exchange on collapse mechanisms and stability of magnetic skyrmions},
  author = {Schrautzer, Hendrik and von Malottki, Stephan and Bessarab, Pavel F. and Heinze, Stefan},
  journal = {Phys. Rev. B},
  volume = {105},
  issue = {1},
  pages = {014414},
  numpages = {21},
  year = {2022},
  month = {Jan},
  publisher = {American Physical Society},
  doi = {10.1103/PhysRevB.105.014414},
  url = {https://link.aps.org/doi/10.1103/PhysRevB.105.014414}
}

@article{Demiroglu2024DFTILDMI,
  title = {{Oscillatory behavior of interlayer Dzyaloshinskii-Moriya interaction by spacer thickness variation}},
  author = {Demiroglu, E. and Hancioglu, K. and Yavuz, I. and Avci, C. O. and Deger, C.},
  journal = {Phys. Rev. B},
  volume = {109},
  issue = {14},
  pages = {144422},
  numpages = {6},
  year = {2024},
  month = {Apr},
  publisher = {American Physical Society},
  doi = {10.1103/PhysRevB.109.144422},
  url = {https://link.aps.org/doi/10.1103/PhysRevB.109.144422}
}

@article{pollard2020zdmi,
  title = {Bloch Chirality Induced by an Interlayer {D}zyaloshinskii-{M}oriya Interaction in Ferromagnetic Multilayers},
  author = {Pollard, Shawn D. and Garlow, Joseph A. and Kim, Kyoung-Whan and Cheng, Shaobo and Cai, Kaiming and Zhu, Yimei and Yang, Hyunsoo},
  journal = {Phys. Rev. Lett.},
  volume = {125},
  issue = {22},
  pages = {227203},
  numpages = {7},
  year = {2020},
  month = {Nov},
  publisher = {American Physical Society},
  doi = {10.1103/PhysRevLett.125.227203},
  url = {https://link.aps.org/doi/10.1103/PhysRevLett.125.227203}
}

@article{Simon2018,
  title = {Magnetism of a {Co} monolayer on {P}t(111) capped by overlayers of $5d$ elements: A spin-model study},
  author = {Simon, E. and R\'ozsa, L. and Palot\'as, K. and Szunyogh, L.},
  journal = {Phys. Rev. B},
  volume = {97},
  issue = {13},
  pages = {134405},
  numpages = {11},
  year = {2018},
  month = {Apr},
  publisher = {American Physical Society},
  doi = {10.1103/PhysRevB.97.134405},
  url = {https://link.aps.org/doi/10.1103/PhysRevB.97.134405}
}

\section*{Acknowledgments}
T.M. acknowledges fruitful discussions with A. Tripathi and A. Hierro Rodriguez. E.Y.V. and T.M. acknowledge financial support provided by the Deutsche Forschungsgemeinschaft (DFG) via Project No. 514141286. L.R. gratefully acknowledges funding by the National Research, Development, and Innovation Office (NRDI) of Hungary under Project Nos. FK142601 and ADVANCED 149745, by the Ministry of Culture and Innovation and the National Research, Development and Innovation Office within the Quantum Information National Laboratory of Hungary (Grant No. 2022-2.1.1-NL-2022-00004), and by the Hungarian Academy of Sciences via a J\'{a}nos Bolyai Research Grant (Grant No. BO/00178/23/11).

\section*{Author contributions}
T.M., L.R., R.W., and E.Y.V. conceived the study. T.M. carried out the calculations, and prepared the figures. L.R. and E.Y.V. supervised the project. T.M., L.R., and E.Y.V. were major contributors in writing of the manuscript. All authors discussed results and reviewed the manuscript.

\section*{Competing interests}
The authors declare no competing financial or non-financial interests.

\section*{Correspondence}
Correspondence and requests for materials should be addressed to Levente R\'{o}zsa or Elena Y. Vedmedenko.


\clearpage
\section*{Supplementary Note 1: Helicity dependence of the twin-skyrmion on the IL-DMI}
\label{sec:HelicityDerivation}
To study the effect of the IL-DMI on the skyrmion helicity analytically, we take the continuum limit $\vec{S}_{i}^{l}\to \vec{s}_{l}(\vec{r})$ for $l\in\{(1),(2)\}$. Similarly to the atomistic Hamiltonian Eq.~(7), the continuum free-energy density can be split into three parts,
\begin{equation}
    \mathcal{E}=\mathcal{E}_1\qty[\vec{s}_1 \qty(\vec{x}) ]+\mathcal{E}_2[\vec{s}_2 \qty(\vec{x}) ]+\mathcal{E}_{\text{inter}}[\vec{s}_1 \qty(\vec{x}),\vec{s}_2 \qty(\vec{x}) ].
     \label{eq:A1}
\end{equation}
The intralayer interactions are contained in $\mathcal{E}_{l}\qty[\vec{s}_{l} \qty(\vec{x}) ]$,
\begin{align}
    \mathcal{E}_{l}\qty[\vec{s}_{l} \qty(\vec{x}) ]=& \mathcal{E}_{l}^{\text{Exch}}+\mathcal{E}_{l}^{\text{DMI}}+\mathcal{E}_{l}^{\text{Zeeman}}\nonumber\\=& \mathcal{A}\qty( \qty(\vec{\nabla}s^x_{l})^2+\qty(\vec{\nabla}s^y_{l})^2+\qty(\vec{\nabla}s^z_{l})^2) \nonumber\\
    +& \mathcal{D}^{\text{IF}}\qty(s^z_{l} \partial_x s^x_{l}-s^x_{l} \partial_x s^z_{l}+s^z_{l} \partial_y s^y_{l}-s^y_{l} \partial_y s^z_{l})\nonumber\\
    -&\mathcal{M}B^{z}s^z_{l},
     \label{eq:A2}
\end{align}
where we assumed that the exchange stiffness $\mathcal{A}$, the N\'{e}el-type DMI constant $\mathcal{D}$, the saturation magnetization $\mathcal{M}$, and the magnetic field $B^{z}$ are the same for the two layers, as in Eq.~(8). The two layers are coupled in the following way:
\begin{align}
    \mathcal{E}_{\text{inter}}=&-\mathcal{J} \qty( s_1^x s_2^x +s_1^y s_2^y+s_1^z s_2^z )  +\mathcal{D}^{z}\cdot\qty(s_1^x s_2^y - s_1^y s_2^x),
     \label{eq:A3}
\end{align}
where $\mathcal{J}$ is the IEC constant and $\mathcal{D}^{z}$ is the $z$ component of the IL-DMI vector. We assume that the other components are zero, $\mathcal{D}^{x}=\mathcal{D}^{y}=0$. Under this assumption, the Euler--Lagrange equations derived from Eq.~\eqref{eq:A1} admit circularly symmetric solutions, as has been extensively studied in the case of a single layer~\cite{Bogdanov1999}. To describe these spin textures, we use polar coordinates in space,
\begin{align}
    \mqty(x \\ y)&=\mqty(\rho \sin \varphi \\ \rho \cos \varphi),\\
    \mqty(\partial_x \\ \partial_y)&=\mqty(\cos \varphi \partial_\rho - \rho^{-1} \sin \varphi \partial_\varphi \\ \sin \varphi \partial_\rho + \rho^{-1}  \cos \varphi \partial_\varphi).
    \label{eq:A5}
\end{align}
In spin space, we transform to spherical coordinates,
\begin{align}
    \mqty(s^x_{l} \\ s^y_{l} \\ s^z_{l})&=\mqty(\sin \Theta_{l} \cos \Phi_{l}\\ \sin \Theta_{l} \sin \Phi_{l}\\ \cos \Theta_{l} ).
    \label{eq:A6}
\end{align}
Equilibrium spin structures are found as the solution of the Euler--Lagrange equations,
\begin{align}
        \frac{\dd}{\dd \rho}\qty(\frac{\partial \mathcal{E}}{\partial (\partial_{\rho}\Theta_{l})})+\frac{1}{\rho}\qty(\frac{\partial \mathcal{E}}{\partial (\partial_{\rho}\Theta_{l})})&=\frac{\partial \mathcal{E}}{\partial \Theta_{l}},\label{eq:A14}\\
        \frac{\dd}{\dd \varphi}\qty(\frac{\partial \mathcal{E}}{\partial (\partial_{\varphi}\Theta_{l})})&=\frac{\partial \mathcal{E}}{\partial \Theta_{l}},\\
        \frac{\dd}{\dd \rho}\qty(\frac{\partial \mathcal{E}}{\partial (\partial_{\rho}\Phi_{l})})+\frac{1}{\rho}\qty(\frac{\partial \mathcal{E}}{\partial (\partial_{\rho}\Phi_{l})})&=\frac{\partial \mathcal{E}}{\partial \Phi_{l}},\\
        \frac{\dd}{\dd \varphi}\qty(\frac{\partial \mathcal{E}}{\partial (\partial_{\varphi}\Phi_{l})})&=\frac{\partial \mathcal{E}}{\partial \Phi_{l}}.\label{eq:A15}
\end{align}
We use the following ansatz for circular skyrmions:
\begin{align}
    \Theta_{l}(\rho,\varphi)&=\Theta_{l}(\rho), \\ \Phi_{l}(\rho,\varphi)&=m_{l}\varphi+\psi_{l} (\rho).
\end{align}
The type of IF-DMI assumed in Eq.~\ref{eq:A2} prefers the vorticity $m_{l}=1$ and the N\'{e}el-type helicity $\psi=0$ in a single layer. The IL-DMI does not change the vorticity, but induces opposite fields in the two layers, and the equilibrium conditions may be simultaneously satisfied for $\psi_{l}(\rho)=\pm\psi(\rho)$. The same material parameters in the two layers also allow for the assumption $\Theta_1(\rho)=\Theta_2(\rho)=\Theta(\rho)$. 
Combining Eqs.~\eqref{eq:A5} and \eqref{eq:A6}, one can rewrite Eqs.~\eqref{eq:A2} and \eqref{eq:A3} in polar coordinates as
\begin{align}
    \mathcal{E}_{l}^{\text{Exch}}&=\mathcal{A} \qty(( \Theta')^2+(\psi')^2\sin^2 \Theta+\frac{\sin^2 \Theta}{\rho^2}),\\
    \mathcal{E}_{l}^{\text{DMI}}&=\mathcal{D}^{\text{IF}} \bigg(\qty( \Theta' + \frac{\sin \Theta \cos \Theta}{\rho}   )\cos \psi\nonumber\\
    &\quad\quad\quad\quad\quad-\psi'\sin\Theta\cos\Theta\sin\psi\bigg)
    ,\\
    \mathcal{E}_{l}^{\text{Zeeman}}&=-\mathcal{M} B^{z}
    \cos \Theta(\rho),
\end{align}
and
\begin{align}
    \mathcal{E}_{\text{inter}}=&-\mathcal{J}^{\text{IL}}\qty(\sin^2\Theta\cos\qty(2\psi) +\cos^2 \Theta)\nonumber\\
    &+\mathcal{D}^z \sin^2 \Theta \sin \qty(2\psi),
    \label{eq:A9}
\end{align}
respectively. $\Theta'$ and $\psi'$ are the radial derivatives $\partial_\rho\Theta(\rho)$ and $\partial_\rho\psi(\rho)$. Only the difference in azimuthal angles $\Phi_1-\Phi_2$ enters Eq.~\eqref{eq:A9}, which simplifies to $\Phi_1-\Phi_2=2\psi(\rho)$. These assumptions simplify Eqs.~\eqref{eq:A14}-\eqref{eq:A15} to two coupled radial differential equations,
\begin{align}
    \Theta''= &- \frac{1}{\rho} \Theta'+ \qty((\psi')^2  +\frac{1}{\rho^2})\sin \Theta \cos \Theta \nonumber\\&-\frac{1}{2\mathcal{A}}\bigg(2\mathcal{D}^{\text{IF}} \sin^2 \Theta\qty(\frac{1}{\rho}\cos \psi - \psi' \sin \psi   ) \nonumber \\&+\sin\Theta\cos\Theta(\mathcal{J}^{\text{IL}} (\cos (2\psi)-1) 
    -\mathcal{D}^z  \sin (2\psi))\nonumber \\&-\mathcal{M} 
    B^{z}\sin \Theta\bigg),\label{eq:A16}\\
    \psi''=&- \frac{1}{\rho} \psi'-\frac{\sin(2\Theta)}{\sin^2 \Theta}\Theta'\psi'-\frac{1}{4\mathcal{A} }\bigg(4\mathcal{D}^{\text{IF}} \sin\psi\Theta'\nonumber\\&-2 (\mathcal{J}^{\text{IL}} \sin(2\psi)+\mathcal{D}^z\cos(2\psi)) \bigg). \label{eq:A17}
\end{align}
Note that for $\psi=\psi'=0$, Eq.~\eqref{eq:A16} simplifies to the radial Euler--Lagrange equation of a single skyrmion~\cite{Bogdanov1999}. However, this assumption does not satisfy Eq.~\eqref{eq:A17} in the presence of an IL-DMI $\mathcal{D}^z$. For a fixed value of $\psi$, the interlayer terms introduce an effective anisotropy term proportional to $\sin \Theta \cos \Theta$. This prefers an in-plane alignment of the spins since in this case they can gain energy from the IL-DMI, thereby it competes with the external field and results in an increase of the skyrmion size.

\begin{figure}
    \centering
    \includegraphics[width=\linewidth]{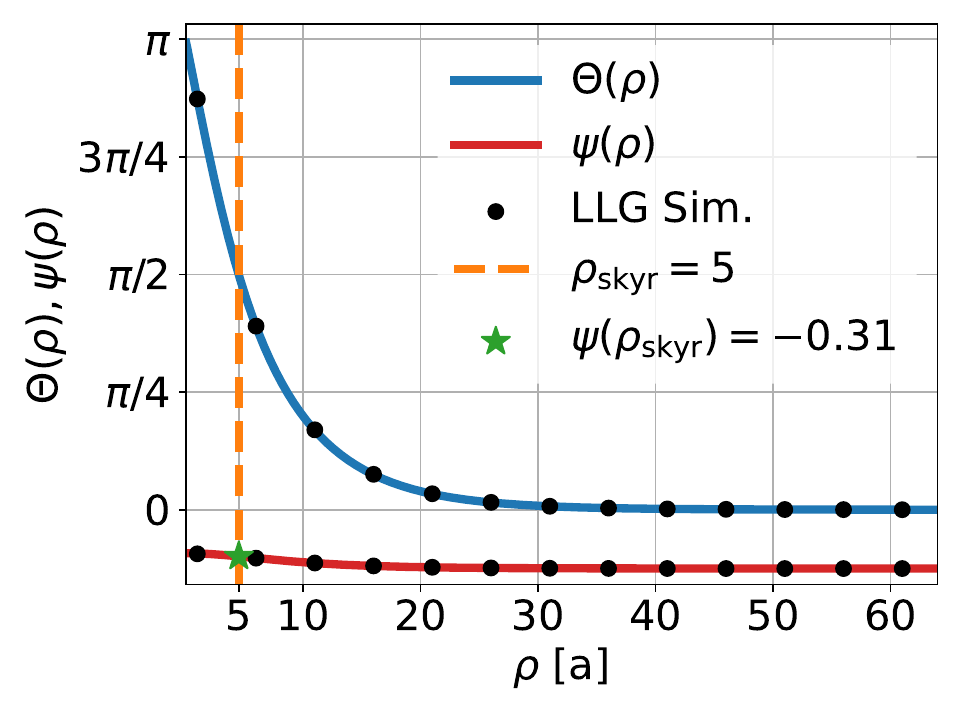}
    \caption{Skyrmion profile from the continuum model for out-of-plane IL-DMI. The solid lines show the numerical solutions of Eqs.~\eqref{eq:A16} and \eqref{eq:A17} under the boundary conditions in Eq.~\eqref{eq:A18}. The black dots are obtained from the atomistic LLG simulations. The highlighted value $\psi(\rho_{\textrm{skyr}})=\frac{1}{2}(\psi_1-\psi_2)=0.311$ shows the helicity at the skyrmion radius $\Theta(\rho_{\textrm{skyr}})=\pi/2$. The system parameters are: $J^{\text{IF}}=\SI{6}{meV}$, $D^{\text{IF}}=\SI{1.5}{meV}$, $B^{z}=\SI{3}{T}$, $J^{\text{IL}}=\SI{1}{meV}$, and $D^{\text{IL}}_z=\SI{1}{meV}$. The continuum parameters are $\mathcal{A}=\SI{3}{meV}/a^{3}$, $\mathcal{D}^{\text{IF}}=\SI{1.5}{meV }/a^2$, $B^z=\SI{3}{T}$, $\mathcal{J}^{\text{IL}}=\SI{1}{meV}/a^{3}$, and $\mathcal{D}^z=\SI{1}{meV}/a^{3}$. The conversion between the atomistic and the continuum parameters follows the expressions given in Ref.~\cite{rozsa2017TempScaling} for the intralayer terms; the interlayer parameters are simply divided by the atomic volume when passing to the continuum limit.}
    \label{fig:skyrprofile}
\end{figure}

We solve Eqs.~\eqref{eq:A16} and \eqref{eq:A17} numerically with the boundary conditions
\begin{equation}
    \Theta(0)=\pi,\;\Theta(\infty)=0, \;\psi'(0)=0, \text{ and } \psi'(\infty)=0.\label{eq:A18}
\end{equation}
We solve the boundary value problem using the implementation of SciPy~\cite{2020SciPy-NMeth}, which is an adaption of the algorithm presented in~\cite{bvp2001}. We compare the numerical solution of Eqs.~\eqref{eq:A16} and~\eqref{eq:A17} with the profile obtained from the atomistic simulations via a solution of the LLG equation in Fig.~\ref{fig:skyrprofile}. The lattice and the continuum models agree well. Since the helicity $\psi$ only depends weakly on the radial coordinate, it may be well described by its value at the skyrmion radius $\Theta(\rho_{\textrm{skyr}})=\pi/2$, which was used for the comparison between different IL-DMI values in Fig.~3 in the main text.

\section*{Supplementary Note 2: Skyrmion Hall angle as a function of helicity}
\label{app:HallAngle}

For circularly symmetric skyrmions, the Hall angle may be calculated analytically from the Thiele equation in Eq.~(3). Using that the dissipation tensor $\mathfrak{D}$ is a unit tensor with magnitude $\mathcal{D}$, and the dependence of the $\mathcal{B}$ tensor on the helicity $\psi$ and the vorticity $m=1$ is given by Eq.~(6), the skyrmion velocity may be expressed as
\begin{equation}
     \begin{pmatrix} v_x \\ v_y\end{pmatrix} = \frac{f^{\text{SOT}}}{\alpha^2\mathcal{D}^2+G^2} \begin{pmatrix} \alpha \mathcal{D} & G \\ -G& \alpha \mathcal{D}\end{pmatrix} \begin{pmatrix}\sin \psi & -\cos \psi \\ \cos \psi & \sin \psi\end{pmatrix}\begin{pmatrix} P_x \\ P_y\end{pmatrix}.
\end{equation}
We omitted $P_z$ because it does not contribute to the force $\vec{F}$.

The skyrmion Hall angle $\theta_{\text{Hall}}$ is defined as the angle between the velocity and the current direction. We connect the current direction to the velocity via $\vec{P}\parallel\vec{j}_{\textrm{el}}\cross\hat{\vec{z}}$, as is usual for the spin--orbit torque, resulting in $\vec{j}_{\textrm{el}}\parallel\mqty(-P_y & P_x)$. For a N\'{e}el-type skyrmion with $\psi=0$, this coincides with the direction of the force $\vec{F}=\mathcal{B}\vec{P}$. 
For circularly symmetric skyrmions, we can assume $\vec{P}=\mqty(1 & 0)$ without loss of generality, resulting in $\vec{j}_{\textrm{el}}\parallel\mqty(0 & 1)$. We obtain the following for the direction of the skyrmion velocity:
\begin{equation}
    \mqty(v_x\\v_y)\parallel \mqty( k\sin\psi+\cos\psi\\ k\cos\psi- \sin\psi ),
\end{equation}
where $k=\frac{\alpha \mathcal{D}}{G}$. The skyrmion Hall angle is given by
\begin{align}
    \theta_{\text{Hall}}^{\text{eff}}
    &=\arctan \qty(\frac{ k\cos\psi- \sin\psi}{k\sin\psi+\cos\psi})-\frac{\pi}{2}\\
    &=\arctan \qty(\frac{ k- \tan\psi}{k\tan\psi+1})-\frac{\pi}{2}\\
    &= \arctan k -\arctan (\tan\psi)-\frac{\pi}{2}\\
    &=\theta_{\text{Hall}}^{\psi=0}-\psi,
    \label{eq:HelicityHall}
\end{align}
where $\theta_{\text{Hall}}^{\psi=0}=\arctan (k)-\frac{\pi}{2}=\arctan \left(1/k\right)=\arctan \left(G/\alpha\mathcal{D}\right)$ is the typical expression given in the literature for the skyrmion Hall angle~\cite{Weienhofer2022Stochastic,jiang2017direct,Xia2020Elliptical}. This implies that the two parts of the twin-skyrmion are driven along different directions by the spin-polarized current due to their different helicities, which results in an instability of the motion for high driving currents. 
\section*{Supplementary Note 3: Skyrmion velocity as a function of skyrmion radius}
\label{app:BTensorSize}
Under the assumption of circularly symmetric skyrmions for out-of-plane IL-DMI, the dependence of the velocity on the skyrmion size may also be calculated. To derive an analytical estimate for the parameters, we assume a linear dependence of $\Theta$ on the radial coordinate $\varrho$ for the skyrmion profile,
\begin{equation}
    \Theta(\rho) = \begin{cases}
        \pi-\nu \cdot \rho, & \rho<\frac{\pi}{\nu}\\
        0, & \rho \geq \frac{\pi}{\nu}
    \end{cases}.
    \label{eq:Ansatz}
\end{equation}
The radius of the skyrmion is defined as
\begin{equation}
    \pi-\nu \cdot R=\frac{\pi}{2} \Rightarrow R =\frac{\pi}{2\nu}.
\end{equation}
The factor $f^{\textrm{SOT}}$ in the $\mathcal{B}$ tensor in Eq.~(6) is given by the integral
\begin{equation}
    f^{\textrm{SOT}}=\frac{\beta_d \mu_{s}\pi}{a^{2}\gamma} \int _0 ^\infty \qty[\sin\Theta \cos \Theta + \rho \frac{\partial \Theta}{\partial \rho}] \dd \rho.
    \label{eq:intB}
\end{equation}
which for the given skyrmion profile may be evaluated as
\begin{align}
    f^{\textrm{SOT}}=\frac{\beta_d \mu_{s}\pi}{a^{2}\gamma}
    \left(-\frac{\pi^2}{2\nu}\right)=-\frac{\beta_d \mu_{s}\pi^{2}}{a^{2}\gamma}\cdot R.
\end{align}
Therefore, the magnitude of the force scales linearly with the size of the skyrmion. 

\begin{figure}
    \centering
    \includegraphics[width=\linewidth]{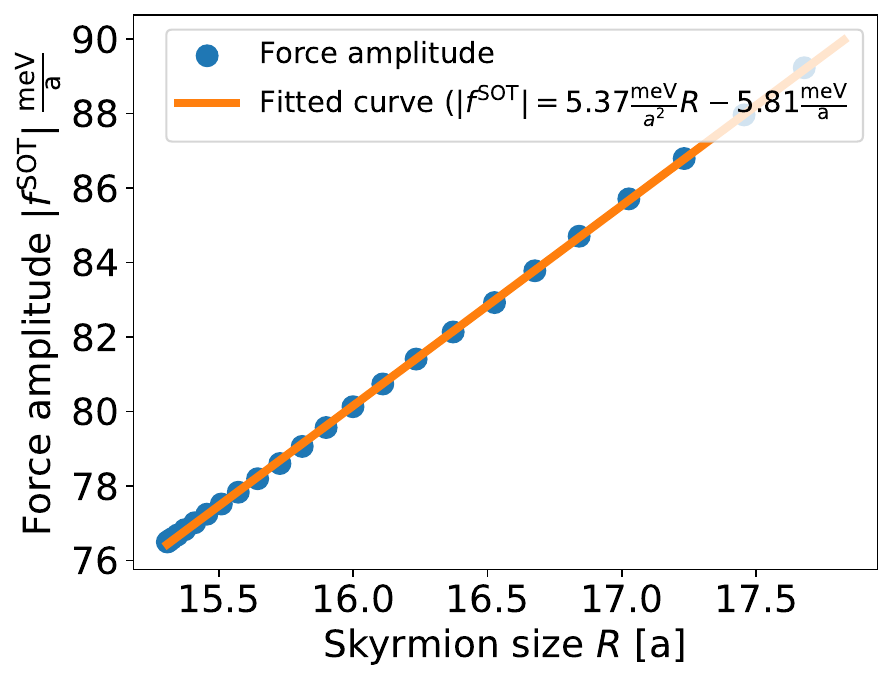}
    \caption{
    Scaling of the force amplitude $f^{\textrm{SOT}}$ with the skyrmion radius $R$. 
    The different skyrmion sizes are generated by changing the IL-DMI $D_z^{\text{IL}}$ from \SI{0.0}{meV} to \SI{0.6}{meV}, using the same values as in Fig.~5 in the main text. The blue points are obtained by numerically integrating Eq.~\eqref{eq:intB} with $\Theta(\rho)$ coming from the atomistic simulations.}
    \label{fig:b_tensor}
\end{figure}

We confirm this linear scaling by plotting the values of $f^{\textrm{SOT}}$ obtained from simulations in Fig.~\ref{fig:b_tensor}. 

The dissipation tensor $\mathfrak{D}$ for circular skyrmions is a unit tensor, with its magnitude given by 
\begin{equation}
    \mathcal{D}=\frac{\mu_{s}}{a^{2}\gamma}\int \qty(\partial_x \vec{s} )^2 \dd ^2 r.
\end{equation}
Performing the coordinate transformation to polar coordinates in real space and spherical coordinates in spin space, one obtains,
\begin{equation}
    \mathcal{D}=\frac{\pi \cdot \mu_{s}}{a^{2}\gamma}\int_0^\infty \qty[\rho\Theta'(\rho)^2  \cos^2 \Theta(\rho)+\frac{1}{\rho}\sin^2 \Theta(\rho)] \dd \rho.
\end{equation}
Using the expression for $\Theta(\rho)$ in Eq.~\eqref{eq:Ansatz}, we obtain
\begin{align}
    \mathcal{D}=&\frac{\pi \cdot \mu_{s}}{a^{2}\gamma}\int_0^{\frac{\pi}{a}} \qty[\frac{1}{\rho}\sin^2 (a\cdot\rho)-\rho \cdot a^2  \cos^2 (a\cdot\rho)] \dd \rho.\\
    =&\frac{\pi \cdot \mu_{s} }{a^{2}\gamma }\int_0^{\pi} \qty[\frac{1}{x}\sin^2 (x)-x \cdot   \cos^2 (x)] \dd x.
\end{align}
This demonstrates that $\mathcal{D}$ is scale independent, i.e., it does not depend on the skyrmion radius if the shape of $\Theta(\rho)$ does not change. 
This can also be observed in the simulations: as $D_z^{\text{IL}}$ increases from \SI{0.0}{meV} to \SI{0.6}{meV}, the value of $\mathcal{D}$ numerically calculated from the atomistic model only decreases by 3.3\% from $\SI{15.3}{meV\cdot s/a^2}$ to $\SI{14.8}{meV\cdot s/a^2}$, while $f^{\textrm{SOT}}$ increases by 18.4\% for the same parameters as in Fig.~5 in the main text.

\section*{Supplementary Note 4: Instability of the parallel magnetic alignment of the two layers in the presence of $D_z^{\textrm{IL}}$}
\label{app:zDMIvsB}
For a sufficiently strong external magnetic field $\vec{B}$, the ground-state magnetic configuration obtained from the Hamiltonian in Eq.~(8) and (9) in the main text is collinear in each layer, $\vec{S}_{i}^{l}=\vec{S}^{l}$. 
This simplifies the Hamiltonian to the following macrospin model:
\begin{equation}
    H=-J^{\textrm{IL}}\vec{S}^{(1)}\cdot \vec{S}^{(2)}+\vec{D}^{\textrm{IL}} \cdot (\vec{S}^{(1)}\times \vec{S}^{(2)})-\mu_s\vec{B}\cdot(\vec{S}^{(1)} + \vec{S}^{(2)}).\label{eq:D1}
\end{equation}
We assume that the IL-DMI vector and the magnetic field point along the $z$ direction, $\vec{D}^{\textrm{IL}}=D^{IL}_z\hat{\vec{z}}$ and $\vec{B}=B\hat{\vec{z}}$. Using the spherical coordinates analogously to Eq.~\eqref{eq:A6}, we can rewrite Eq.~\eqref{eq:D1} as
\begin{align}
    H=&-J^{\textrm{IL}}(\sin^2 \Theta \cos (\Phi_1-\Phi_2) +\cos ^2 \Theta )\nonumber\\&+D^{IL}_z \sin^2\Theta\sin (\Phi_2-\Phi_1)\nonumber\\&-2\mu_s B\cos \Theta.\label{eq:D2}
\end{align}
Here, we assume that the polar angle is the same for both spins, $\Theta_1=\Theta_2=\Theta$, which follows from the symmetry of the two layers. We introduce the variable $\delta=\Phi_1-\Phi_2$. To find the spin configuration with minimal energy, we require
\begin{align}
    \partial_{\Theta} H = J^{\textrm{IL}}\sin 2\Theta\qty(1- \cos \delta -\frac{D^{IL}_z}{J^{\textrm{IL}}}\sin\delta)\nonumber\\+2\mu_sB\sin \Theta=0,
    \label{eq:D3}
\end{align}
\begin{align}
    \partial_\delta H = (J^{\textrm{IL}} \sin \delta -D^{IL}_z  \cos \delta)\sin^2\Theta=0.
    \label{eq:D4}
\end{align}
Both equations are satisfied for $\sin\Theta=0$, i.e., when the spins are aligned along the $z$ direction. Solutions with $\sin \Theta\neq0$ can be found by transforming Eq.~\eqref{eq:D4} to
\begin{equation}
    \delta = \arctan \frac{D^{IL}_z}{J^{\textrm{IL}}}.\label{eq:D5}
\end{equation}
Inserting Eq.~\eqref{eq:D5} into Eq.~\eqref{eq:D3} gives
\begin{equation}
    2 \cos \Theta \sin \Theta ( \sqrt{{J^{\textrm{IL}}}^2+{D^{IL}_z}^2} -J^{\textrm{IL}})=2\mu_sB\sin \Theta,
\end{equation}
which may be solved for $\Theta$ as
\begin{equation}
    \Theta =\arccos(\frac{\mu_sB}{\sqrt{{J^{\textrm{IL}}}^2+{D^{IL}_z}^2} -J}). \label{eq:D7}
\end{equation}
This expression can only be evaluated if the argument of the $\arccos$ function is in $[-1,1]$. This condition may be reformulated as
\begin{equation}
    \left|D^{IL}_z(J^{\textrm{IL}},B)\right|\ge\sqrt{(\mu_sB\pm {J^{\textrm{IL}})}^2-(J^{\textrm{IL}})^2}.
    \label{eq:D8}
\end{equation}
It is straightforward to show that if this solution exists, it has a lower energy than the $\sin\Theta=0$ solution. In this case, the spins deviate from the direction of the external field since they gain more energy from the IL-DMI by forming a finite angle in the $xy$ plane. For the parameters $J^{\textrm{IL}}=\SI{1}{meV},\mu_{s}=3\mu_{\textrm{B}}$ and $B=\SI{3}{T}$ used as an example in the main text, this transition occurs for $\left|D^{IL}\right|=\SI{1.15}{meV}$.
\end{document}